\documentstyle[12pt]{article} 

\def\baselinestretch{2}

\def\setstretch#1{\renewcommand{\baselinestretch}{#1}}

\def\singlespace{\def\baselinestretch{1}\@normalsize}

\def\@setsize#1#2#3#4{\@nomath#1\let\@currsize#1\baselineskip
   #2\baselineskip\baselinestretch\baselineskip
   \setbox\strutbox\hbox{\vrule height.7\baselineskip
      depth.3\baselineskip width\z@}
   \normalbaselineskip\baselineskip#3#4}

\skip\footins 20pt plus4pt minus4pt

\def\@xfloat#1[#2]{\ifhmode \@bsphack\@floatpenalty -\@Mii\else
   \@floatpenalty-\@Miii\fi\def\@captype{#1}\ifinner
      \@parmoderr\@floatpenalty\z@
    \else\@next\@currbox\@freelist{\@tempcnta\csname ftype@#1\endcsname
       \multiply\@tempcnta\@xxxii\advance\@tempcnta\sixt@@n
       \@tfor \@tempa :=#2\do
                        {\if\@tempa h\advance\@tempcnta \@ne\fi
                         \if\@tempa t\advance\@tempcnta \tw@\fi
                         \if\@tempa b\advance\@tempcnta 4\relax\fi
                         \if\@tempa p\advance\@tempcnta 8\relax\fi
         }\global\count\@currbox\@tempcnta}\@fltovf\fi
    \global\setbox\@currbox\vbox\bgroup 
    \def\baselinestretch{1}\@normalsize
    \boxmaxdepth\z@
    \hsize\columnwidth \@parboxrestore}
\long\def\@footnotetext#1{\insert\footins{\def\baselinestretch{1}\footnotesize
    \interlinepenalty\interfootnotelinepenalty 
    \splittopskip\footnotesep
    \splitmaxdepth \dp\strutbox \floatingpenalty \@MM
    \hsize\columnwidth \@parboxrestore
   \edef\@currentlabel{\csname p@footnote\endcsname\@thefnmark}\@makefntext
    {\rule{\z@}{\footnotesep}\ignorespaces
      #1\strut}}}

\input amssym.def
\input epsf.tex
\let\nopictures=N
\newlength{\listindent}
\newlength{\listsep}
\newenvironment{biblist}{\begin{list}{ }{\setlength{\listparindent}{-\listindent}
\setlength{\leftmargin}{\listindent}\itemindent =
\listparindent \parsep=\listsep}\item}{\end{list}}
\newcommand{\vw}{\vec{w}}
\newcommand{\vv}{\vec{v}}
\newcommand{\vwo}{w_{\cal O}}
\newcommand{\vxi}{\vec{\xi}}
\newcommand{\vxio}{\xi_{\cal O}}
\newcommand{\bw}{\mbox{\boldmath $w$}}
\newcommand{\bxi}{\mbox{\boldmath $\xi$}}
\newcommand{\bv}{\mbox{\boldmath $v$}}
\newcommand{\vr}{\vec{r}}

\newcommand{\etal}{{\sl et al.}}

\newcommand{\BO}[2]{{\sl Biological~Cybernetics},\ {\bf #1},\ #2}

\newcommand{\PRSL}[2]{{\sl Procceedings of the Royal Society of London},\ {\bf
#1},\ #2} 
\newcommand{\NAS}[2]{{\sl Proceedings of the National Academy of Science USA},\
{\bf #1},\ #2}  
\newcommand{\JN}[2]{{\sl Journal of Neuroscience},\ {\bf #1},\ #2}

\begin{document}

\begin{flushright}
April, 1995
\end{flushright}

\vfill
\begin{center}

	 {{\large\bf Spontaneous symmetry breaking and the formation of columnar
structures in the primary visual cortex II}\\ --- Local organization of
orientation modules --- 
   }  

 \vspace{.6in}

 \normalsize 	
{\bf Kengo  Yamagishi} 
															  
 \vspace{.2in}
Lab for Neural Modeling\\ 
email: {\tt k-yama@hep-th.phys.s.u-tokyo.ac.jp}

 \vfill

\bigskip\bigskip
\vfill

 {\bf                     
     ABSTRACT
 }

 \mbox{}

 \parbox{5.5in}
 {\hspace{.2in}Self-organization of orientation-wheels observed in the
visual cortex is discussed from the view point of topology.  We argue in a
generalized model of Kohonen's feature mappings that the existence of the
orientation-wheels is a consequence of Riemann-Hurwitz formula from
topology.  In the same line, we estimate partition function of the model,
and show that regardless of the total number $N$ of the orientation-modules
per hypercolumn the modules are self-organized, without fine-tuning
of parameters, into 
definite number of orientation-wheels per hypercolumn if $N$ is large.  
 }

\end{center}

\bigskip\bigskip


\vfill

\newpage
\parindent.3in

\section{Introduction} 

Among various columns observed in cortical surface, orientation columns in
primary visual area V1 exhibit one of the most attractive
structural organization.  Here expanding our previous study (Yamagishi, 1994,
referred to as I hereafter), we would like to investigate the mechanism of the
self-organization of these orientation columns from the view point of
spontaneous symmetry breaking.  In so doing we hope that we will get deeper
understanding of the role of these columns not only in the context of
self-organization but also in the information  processing problem in visual
cortex. 

Since the first discovery by Hubel and Wiesel (1977 for a survey), these
distributed localized area of iso-orientation preference have 
attracted people's attention. In their earlier model (see Figure 1),
based on their experimental data, although they are not conclusive, the columns
are considered (proposed) as some form of slabs embedded between
ocular dominance boundary walls. Each slab is assigned detectors with specific
orientation preference.  However, in the recent experiments, as more precise
data have become available, more detailed structure of this organization is
becoming to emerge. 

\medskip
\let\picnaturalsize=N
\def\picsize{2.5in}
\def\picfilename{figure1.eps}
\ifx\nopictures Y\else{\ifx\epsfloaded Y\else\relax \fi
\let\epsfloaded=Y
\centerline{\ifx\picnaturalsize N\epsfxsize \picsize\fi
\epsfbox{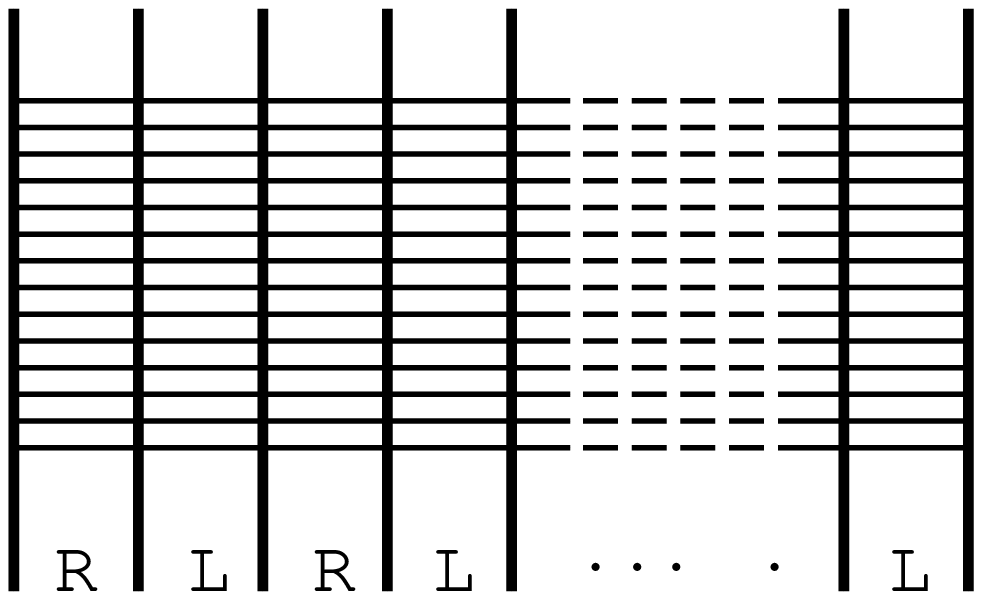}}}\fi

\medskip
In one of the most recent data Blasdel (1992; see also Obermayer \etal, 1993)
showed in higher precision that the orientation columns are basically organized
according to two classes of structural basis: point singularities and
fractures.  The general feature near the ocular dominance boundary wall still
remain the same as in the previous results; the orientation and ocular dominance
walls intersect in right angles.  It has also been shown that these structural
bases --- singular points and/or fractures --- typically stay between the ocular
dominance walls with the average number $(3\sim 4)$ per hypercolumn. This
structure constitutes a unit module (approximately 1mm$\times$1mm square region
on the cortical surface in the case of monkeys) that contains a right set of
feature-detector cells of color-, \hbox{orientation-,}  spatial
frequency-information, etc. for an associated segment of visual field
(corresponding to $0.8^{\circ}\times0.8^{\circ}$ near fovea region).

The purpose of our study is first to consider the precise meaning of modular
organization where the feature space consists of  continuous (or infinite)
degrees of freedom like in this case.  In principle, we would need infinite
number of column-modules if we enforce modular organization \`a la Hubel and
Wiesel in full precision.  At the same time we do not know the minimum
resolution of orientation angles, or even whether such concept at the
orientation-column level is meaningful or not.  In the next section after
reviewing some morphology of this columnar organization, we propose a
``statistical definition of orientation modules" to address this. There the
module boundaries are not defined as definite objects as in classical picture of
Hubel and Wiesel, rather they are defined only statistically.

Quantitative analysis of this scheme is made in \S\S3-4.  We use
generalized Kohonen's feature mapping algorithm (Kohonen, 1982, 1989) to check
this idea.  Models of the same class have been used before for the ocular
dominance columns (Obermayer \etal, 1992, 1990; Yamagishi 1994).  In our model
(I) we added an extra one-dimensional (feature) space to the ordinary Kohonen
that is frequently used in two-dimensional self-organization problem on
cortical surface, and analyzed its continuum limit with respect to spacing
between neurons.  Introducing a certain regularization method to handle rather
singular ``winner-takes-all" mechanism, we derived that the lowest stable state
is a solitonic kink-anti-kink state corresponding to the ocular dominance
domain-walls.  Here we follow basically the same strategy.  Added feature space,
however, is not a line interval $[-1,1]$ but a compact space $S^1$ including its
interior that represents the orientation degrees of freedom with each
probability indicated by the radial distance from the center of the $S^1$.  [The
center of this feature space represents ``no orientation preference".]  In order
to derive a global configuration of the orientation columns, we would need a
full fledged system that includes both ocular and orientation degrees of
freedom.  In this paper, however, we would like to address more fundamental
questions: why the orientation singularities appear, and how these are related
to the (statistical definition of) orientation modules proposed in this
paper.  For the latter purpose only, we do not need feature space terms related
to the ocular dominance columns.

We restrict our model in one hypercolumn area only. [We would like to address
global organization problem in future publications.] The solitonic states in
this case correspond basically to harmonic maps with respect to retinotopic
coordinates.  In two-dimensions (and in a simple topology) they are just linear
combinations of holomorphic and anti-holomorphic mappings.  
As an analytic continuation from global topographic mapping of retina to
visual cortex, which is roughly approximated by a holomorphic (and
anti-holomorphic) mapping, we choose holomorphic mappings to describe
orientation columns in one ocular domain of the hypercolumn (and
anti-holomorphic ones for the other ocular domain).  In modular organization
these mappings are basically one-to-many.  (The image in each orientation
column is a replication of the original one mapped from retina onto
hypercolumn region in the layer 4C.)  Quite generally in a holomorphic mapping
from one compact space to another compact space,  this global information
(multiplicity or ``mapping degree") gives rise to some restriction on the
local structure of the mapping, i.e., existence and number of singularities. 
We will show that the orientation singularities correspond to these
singularities.  We would like to explain this point in detail in \S3.  

In \S4, we derive statistical distribution function (partition function) of the
above mappings in the path integral formalism. (See e.g. our previous work
I.)  The result is written in a summation with respect to the mapping
degree and the number of singularities.  We will see that this precisely
corresponds to our statistical definition of the orientation modules. This is
because the size (accordingly the iso-orientation width) of the module changes
according to the selection of the mapping degrees from the layer 4C to the 
layers 2+3; the mapping degree is large if the iso-orientation width is narrow,
and is small in the converse case.  The partition function contains all these
contributions in the statistical sum.  

One interesting result in our calculation is that  this partition function
takes maximum value at \underline{finite number of singularities
regardless of the mapping degree if the} \underline{latter is large enough}. 
This implies that in average there exist  stable, definite number of
orientation wheels in the hypercolumn anywhere on the cortical surface,
regardless of the size of the orientation modules.  That agrees with what
actually  observed on the cortical surface.

Finally we discuss some future issues in \S5.

\section{Definition of orientation modules}
In this section we present our precise definition of orientation modules,
starting with their fundamental anatomical descriptions on which our theory is
based.

Our main focus in this paper is  the orientation columns in layers 2+3
observed in macaque visual cortex\footnote[2]{Localized areas of orientation
preference are also observed in other layers (5, 6) and visual area (e.g. V2
(area 18) Bonhoeffer \etal, 1993).  However, we would like to restrict our 
theory only to layers 2+3 for the sake of precaution.}.  Major inputs to
neurons in these orientation columns  come from layer 4\footnote[3]{Some from 
4A, 4B, and some directly from 4C$\alpha$, 4C$\beta$ and LGN.}.  Typical
dendrites of the orientation detector neurons extend $200\sim  500\mu$m, that 
is about the scale of the ocular dominance width, in all directions almost
uniformly. Lateral spread of the interlaminar axons from layer 4C is  300$\mu$m
$\sim 400\mu$m right before the dendritic input into cells in layers 2+3. Thus
precise topographic mapping from retina established in the layer 4C is
maintained there only in the scale larger than hypercolumn size (Fitzpatric
\etal,  1985).   Some like-orientation columns in neighboring hypercolumns are
interconnected like clusters via rather long (a few mm up to several mm)
collateral axonal arbors from each orientation cell across the ocular walls
(Gilbert and Wiesel,  1983, 1989). Some cross-orientation (inhibitory)
interactions between such clusters are also observed (Matsubara \etal, 1987;
McGuire \etal, 1991).

The receptive field profile of each constituent (simple) orientation detector
cell ranges typically around $0.5^\circ\!\times0.5^\circ\sim
1.5^\circ\!\times1.5^\circ$ at central visual field.  That has been compared to
2D Gabor filters, and with an appropriate adjustment of parameters a good
agreement is demonstrated (Jones and Palmer,  1987).\footnote[1]{These results
are not conclusive. Other possibilities are also known (e.g. Hawken and Parker,
 1987).}  

In recent high precision experimental data (Blasdel, 1992), observed
iso-orientation domains are organized based on two structural centers: point
singularities and fractures. Distribution of the iso-orientation domains around
the singular point is either increasing or decreasing with respect to the
orientation preference angles. In average, those clock-wise or anti-clock wise
orientation centers distribute evenly in every hypercolumn.  

Statistically among clock wise (or anti-clock wise) orientation centers the
position of an orientation preference within the orientation wheels seems to be
at random, namely, in average one orientation preference (say, 3:00 direction 
in visual field) occurs at any angle in the orientation wheels  on the cortical
surface, although the orientation-preference changes monotonically
increasing or decreasing within one wheel (with a certain increment depending
on the experimental set-up).  It should also be emphasized that the
iso-orientation domain is not separated into more than two pieces within  one
wheel --- it occurs only once. This plays important role later.

This monotonic change of orientation preference is interrupted and cut (or
stop-ped) by ``fractures" where a portion of iso-orientation domains from other
wheels intersects.

The shape of the iso-orientation domains in each wheel does not seem to be
definitive.  It tends to be  elongated in the direction of the
radial direction from the singularity on the cortical surface, but is not
definitive.  It  involves some artifact, i.e., how one defines iso-orientation
width. There is almost no correlation between those shapes and preferred
orientation directions of the cells as well as positions of the iso-orientation
domains in the wheel.

From these observations we can safely conclude the followings. Firstly, the
filtering mechanism of \underline{each} orientation information, via e.g.
2D-Gabor filter-like structures, seems to be independent of the 
self-organization mechanism (since the shape of the iso-orientation domain is
rather arbitrary and its position within orientation wheels does not depend on
the orientation preference of the visual field).

Secondly, it  looks quite unlikely that total monotonic change of 
iso-orientation preferences ranging over 180$^\circ$ in one orientation wheel
plays any essential role in the orientation detection (since the fractures can
interrupt it arbitrarily).

Therefore, this self-organization is something that only assembles
together the cells of similar functionality \underline{semi-locally}   but with
\underline{no definite global order}.  The existence of the
singularities itself looks mere consequence of this organization mechanism
rather than some immediate necessity for information processing strategy of
orientations.  In contrast, the monotonic change (either just increasing or
decreasing, but not like increasing and then decreasing) of the orientation
preference in the semi-local region seems to play important role.

Now, based upon these observations, in the remainder of this section we would
like to present our definition  of the ``modular organization" in the case at
hand, where feature space consists of some continuous degrees of freedom.  In
the original idea of Hubel and Wiesel, each orientation slab, the ``module" of
the orientation detection, was assumed to receive replicated (but somewhat
distorted) image projected over the hypercolumn size region where it resides,
and each of the slabs was assigned detectors with a specific  orientation
preference.  Then those orientation slabs associated to various possible 
orientation directions were supposed to fill up the entire hypercolumn region
as in Figure 1.  In this way, 3-dimensional information (i.e., a totality of
``orientations at every image points") was considered to be processed. 
However, the ``orientation" has full continuous (infinite) degrees of freedom. 
If we enforce this scheme in full precision, we need infinite number of
orientation columns and nerve cells.

However, we also know the following fact. If one tries to
detect an orientation precisely, one has to sacrifice positional
precision that is parallel to that orientation direction (Daugmann,  1985). This
holds not only in each detector cell level but in general.   Hence in the limit
of the finest precision of the orientation the projected image of the 
hypercolumn eventually reduces to one-dimensional degrees of freedom
perpendicular to the orientation direction.  Namely, in the detectable
information space, the  information is not fully 3-dimensional one.  We need a
certain trade-off between the orientation  and the positional precision
parallel to it.

Taking into account this, we consider here the information-processing
scheme and its associated modular organization as in the following.  ``The
module of iso-orientation (and its boundary)  \underline{in the classical sense
of Hubel and Wiesel}\footnote[3]{Here the ``global order" of the
iso-orientation slabs as seen in Figure 1 is not the point of our attention.} 
is not definite".  Namely, we propose that the extent of the module virtually
changes according to the image one is concentrating on.  If one tends to focus
on (via higher cortical function) information about the sharpness of an
orientation direction of an edge rather than whose position (e.g. case of a
long edge), then only the outputs from those detector cells that fall into
quite narrow  range of orientation are \underline{selected} as if those cells
constitute one ``orientation module" associated with that orientation.  In that
case the ``orientation module" encompasses a very thin domain of iso-orientation
region.  On the contrary in the case one tends to extract informations about
orientation with moderate precision at a special, precise point  (e.g. very
short edge --- say, less than a quarter degree $\approx$ 1mm@25cm-distance
sight) in a visual field, then the informations from  moderately narrow range
of orientation preferences are \underline{selected} as if those orientation
cells constitute one orientation module.  Namely, the orientation module is
dynamically configured, and there is no definite boundary between  orientation
modules.   

State it differently, here the size of the orientation module is statistically
defined.  There is no definite (or classical) modular organization in the sense
of Hubel and Wiesel.  

To achieve this multiple projection scheme, almost all the orientation detector
cells have to receive inputs from all axonal arbors projected from the layer 4
 within the same hypercolumn that they belong to.  The results from observation
on axon morphology, as we briefly sketched before about layers 4C and its
superficial laminae, are not inconsistent with this hypothesis. 

The unexpected support for this scheme actually comes from the real
organization of the orientation columns.  The \underline{necessary} 
organizational structure to attain this scheme is firstly the semi-local
continuity of the orientation preference in a size no larger than, say, 30
degrees of orientation width.  We do not need global or $180^\circ$ continuity. 
But most important, among other things, is a forgetful, simple fact that the
iso-orientation patch should not be separated into more than two pieces in one
orientation wheel, namely, the preference angle should not change in such a
fashion that it increases and then decreases continuously.   At least from one
fracture to the next fracture the preference angle has to change monotonically
within one orientation wheel, either increasing or just decreasing, otherwise
our scheme does not work.

This scheme itself does not exclude global order of orientation domains, e.g.,
the organization shown  in Figure 1.  But we do not need such high global order
--- semi-local order is enough to attain this scheme, and that is more plausible
if the cells are self-organized on a non-genetic basis.

\section{Role of holomorphicity}

In this section, we
would like to present our model to describe local-organizations and formulations
of our orientation modules.  We use a model basically  of the same class as
in our previous work I, i.e., a generalization of Kohonen's self-organizing
feature maps.  To establish notations, let us briefly recall its formulations.

\subsection{The model}
Let $\vw(\vr,\tau)$ be a 2-component vector-valued function defined on every
cortical point $\vr$ at time step $\tau$.  We assume that each $\vw(\vr,\tau)$
represent a point in retina's field $E$.  Then the set of vectors
$\{\vw(\vr,\tau)| \vr\in F\equiv{\rm a\ set\ of\ cortical\ points}\}$ defines
neural connections between the retinal surface and the visual cortex, and gives
rise to a mapping of retinal image onto the visual cortex.  
Here, in addition, we consider a
complex-valued function $\vwo(\vr,\tau)$ (where $|\vwo(\vr,\tau)|\le 1$) that
belongs to a hypothetical orientation feature-space $D^2$, a flat disc of radius
1 with its boundary circle $S^1$.  We can think of $\vwo(\vr,\tau)\in D^2$ as
indicating a preference of the  orientation detector cell placed at a cortical
point $\vr$, whose preferred orientation angle is directed to
arg[$\vwo(\vr,\tau)$] with probability $|\vwo(\vr,\tau)|$.  (Namely, the cell
has a polarization $|\vwo(\vr,\tau)|$ in the preferred orientation
arg[$\vwo(\vr,\tau)$].  We assume no orientation preference if
$|\vwo(\vr,\tau)|=0$.)  Then the pair $(\vw(\vr,\tau),\vwo(\vr,\tau))$
represents a retinal information mapped to the cortical point $\vr$ together
with the orientation-preference at that point.

We update this system using the following learning rule.  We start with a random
configuration of $\bw=(\vw(\vr,\tau),\vwo(\vr,\tau))$ at the
time step $\tau=0$, and repeat the two-step process stated below upon every
presentation of the data $\bxi=(\vec{\xi},\xi_{\cal O})\in E\otimes S^1$.  
($\xi_{\cal O}$ is a complex number of modulus 1 specified later.)   The first
step is a so-called ``winner-takes-all":  
\begin{description} 
\item[{[UPDT:1]}]  Select the winning point $\vec{r}^*$ on the cortical surface
where the retinal vector $\bw^*\equiv(\vw(\vr^*,\tau),\vwo(\vr^*,\tau))$ is
closest to the data $\bxi=(\vec{\xi},\xi_{\cal O})$ presented at the time $\tau$.
\end{description}
Here the (distance)$^2$ between two points $\bw_a, \bw_b$ in the feature-data
space is defined as
\begin{equation}
d(\bw_a,\bw_b)\equiv||\vw_a -\vw_b||^2+\sigma_0^2 h({w_{\cal O}}_a,{w_{\cal
O}}_b). 
\end{equation} 
The second term that indicates a perturbative effect from the ordinary Kohonen
becomes significant only in the scale smaller than $\sigma_0$.  We take
$\sigma_0$ the typical size of orientation columns, which is far smaller than
retinotopic organization ($\sim$ scale of the visual area V1).  We also assume
expansion:
\begin{equation}
  h({w_{\cal O}}_a,{w_{\cal O}}_b)=g_1|{w_{\cal O}}_a-{w_{\cal O}}_b|^2+g_2 
  |{w_{\cal O}}_a-{w_{\cal O}}_b|^4+\cdots,
\end{equation}
where $g_1(\sim{\cal O}(1))\gg |g_2|\gg \dots$ are numerical constants taking
real values. As we see later, this choice of $h({w_{\cal O}}_a,{w_{\cal O}}_b)$ 
respects a chiral symmetry of orientation wheels at global scale, namely, the
clock-wise and anti-clock-wise wheels appear evenly in global average.  

The second step is the updating rule for the vector configuration
$\bw\equiv( w_i(\vr,\tau))=(\vw(\vr,\tau),\vwo(\vr,\tau))$  following 
\begin{description} 
\item[{[UPDT:2]}]  $$ <\Delta
w_i(\vr,\tau)P(\bxi,\tau)>_\xi=-\epsilon(\tau)\ \delta{\cal E}[\bw]/\delta
w_i(\vr,\tau),
$$ 
$$
  \Delta w_i(\vr,\tau)\equiv w_i(\vr,\tau+1)-w_i(\vr,\tau).
$$
\end{description}
Here $<\cdots P(\bxi,\tau)>_\xi\equiv\sum_{\mbox{\boldmath\scriptsize
$\xi$}}\cdots P(\bxi,\tau)$ stands for an average over the given probability
function $P(\bxi,\tau)$ of the presented data $\bxi$, $\epsilon(\tau)\ (>0)$ is
the certain (decreasing) learning coefficient $\epsilon(\tau)\downarrow 0$ as
$\tau\rightarrow\infty$, and ${\cal E}[\bw]$ is an energy function that we
specify in the following.  This process stops if the system reaches to a
stationary point $\delta{\cal E}[\bw]/\delta w_i(\vr,\tau)=0$.
 
The energy function we employ is a perturbed one from what frequently used in
the retinotopic self-organization problem.  Let
$\Lambda_{\vr,\vr^*}=\lambda^{-1}\exp(-||\vr-\vr^*||^2/2\sigma_0^2)$ be a
lateral correlation that signifies a short excitatory signal from the winning
point $\vr^*$ with $\lambda$ being a positive numerical constant, and let a
data-profile ${\cal F}_{\vr^*}[\bw]$ at that point be defined as in 
\begin{equation}
{\cal F}_{\vr^*}[\bw]=\left\{ \bxi\equiv(\vec{\xi},\xi_{\cal O})\in E\otimes 
S^1\ |\  \forall\vr\in F,\ d(\bxi,\bw(\vr^*,\tau))\leq
d(\bxi,\bw(\vr,\tau))\right\}.  
\end{equation}
The latter corresponds to a set of data which  a cortical cell at the point
$\vr^*$ can handle under the given configuration of $\bw$.  Using these
quantities, the energy function ${\cal E}[\bw]$ is written as:
\begin{equation}
    {\cal E}[\vw]=\frac{1}{2}\sum_{\vr,\vr^*}\Lambda_{\vr,\vr^*}
 \sum_{\mbox{\boldmath\scriptsize$\xi$}\in{\cal
 F}_{\vr^*}[\mbox{\boldmath\scriptsize $w$}]}P(\bxi,\tau) \left(\|\
  \vec{\xi}-\vw(\vr,\tau)\ \|^2+ 
  \sigma_0^2 h(\xi_{\cal O},w_{\cal O}(\vr,\tau))\right).
\label{energy} 
\end{equation}
As we noted in our previous work I, retaining only $g_1$-term
($g_2=g_3=\cdots=0$) in $h(\xi_{\cal O},w_{\cal O}(\vr,\tau))$ reduces this
system to the standard Kohonen for the mapping of feature space $E\otimes D^2$
onto the cortical surface $F$.

We consider the time-development of this system under presentation of a data
set $\{ \vxi,\xi_{\cal O}\}$ with the probability $P(\vxi,\xi_{\cal
O};\tau)=P(\vxi)$, namely, we use totally random, homogeneous inputs with 
respect to the orientation preference.  We also restrict our orientation data
$\xi_{\cal O}={\rm e}^{{\rm i}\pi k/N}(k=0,1,\dots,N-1)$ for some positive
integer $N$.  This means in particular that each of our input data is always
100\% polarized to one of finite possible orientation directions.

The local symmetry concerned here is unitary group U(1)
\begin{equation}
w_{\cal O}(\vr)\rightarrow{\rm e}^{{\rm i}\gamma(\vr)}\vwo(\vr),\qquad
\vxio\rightarrow{\rm e}^{{\rm i}\gamma(\vr)}\vxio
\label{u1}
\end{equation}
instead of ${\Bbb Z}_2$ appeared in our previous work I.  In this equation  
$\gamma(\vr)$ is an arbitrary real function of $\vr$.  Our energy function
(\ref{energy}) is easily seen to be form-invariant under the transformation
(\ref{u1}).

\subsection{The picture for the ground state}
First we would like to recall results from retinotopy problem derived without
our perturbative term $\sigma_0^2 h(\xi_{\cal O},w_{\cal O}(\vr,\tau))$.  For
this purpose it is convenient to go over to continuum limit of the model  in the
final convergent phase.  We can assume the lateral correlation width
$\sigma_0$ very small.  Then the equation for a stationary state
$\delta{\cal E}[\bw]/\delta w_i=0$ can be approximated as follows. 

At the final convergent stage, the set ${\cal F}_{\vr^*}[\bw]$ is almost 
uniquely given by one point set  ${\cal F}_{\vr^*}[\bw]=\{
\vxi=\vw(\vr^*,\tau)\}$.  Hence we have
$$
0=\frac{\delta{\cal E}[\bw]}{\delta w_i(\vr)}=\int
\Lambda_{\vr,\vr^*}\left(w_i(\vr)-w_i(\vr^*)\right) P(\vw(\vr^*))\ {\rm
d}^2\vw(\vr^*).  
$$
Writing $\vr^*=\vr+\vec{\epsilon}$, and taking into account the Gaussian factor
$\Lambda_{\vr,\vr^*}$ of small width $\sigma_0$, we obtain in the limit 
$\sigma_0\rightarrow0$
\begin{equation}
  0=\partial^2w_i(\vr) +2\left\{P(\vw(\vr))\cdot J(\vw(\vr))\right\}^{-1}
\partial_\rho\! \left\{P(\vw(\vr))\cdot J(\vw(\vr))\right\} \partial_\rho
w_i(\vr).
\label{station}
\end{equation}
Here $J(\vw(\vr))$ is a Jacobian
$$
J(\vw(\vr))=\varepsilon^{\alpha\beta}\partial_\alpha w^1(\vr)\ \partial_\beta
w^2(\vr) 
$$
with anti-symmetric symbol $\varepsilon^{12}=-\varepsilon^{21}=1$, arising
from the change of integration variables ${\rm d}^2\vw(\vr^*)\rightarrow{\rm d}^2
\vec{\epsilon} $.  The stationary state has to satisfy the equation 
(\ref{station}).  

It is pointed out (Ritter and Schulten, 1986) that for the retinotopic
organization an appropriate solution is given  in terms of the harmonic form
satisfying 
\begin{equation}
     \partial^2w_i(\vr)=0.
\label{harmonic}
\end{equation}
In fact given a probability distribution on the retinal data
\begin{equation}
   P(\vxi(\vr))\propto J^{-1}(\vw(\vr))\left|_{\vw(\vr)=\vxi}\right.,
\label{prob}
\end{equation}
one can get a logarithmic mapping relation that solves the equations
(\ref{station}), (\ref{harmonic}),
\begin{equation}
     r^1+{\rm i}r^2=\log (w^1(\vr) \pm {\rm i}w^2(\vr) +\alpha),
\label{retino} 
\end{equation}
between the visual cortex and the retinal surfaces.  This mapping, from half
disk region of $\vw$ to upper half plane of $\vr$,  gives a crude
approximation to the observed results.  In equation (\ref{retino}) the choice of
the sign $\pm$ depends on which ocular (left or right eye) information one
 considering, and $\alpha$ is the constant of length scale of fovea region. 
We presume that the probability distribution (\ref{prob}) is provided via
function $\vw(\vr)$ satisfying (\ref{retino}) at the start of the iteration of
the learning steps.

Next we consider the case with the perturbative term $\sigma_0 h(\xi_{\cal
O},w_{\cal O}(\vr,\tau))$ included, and restrict our attention to one
hypercolumn area only --- that is about 5 to 10 times larger than the scale of
the orientation columns $\sigma_0$.  Then we can still apply the similar
approximation method as we used in deriving the  result (\ref{station}) to the
energy function (\ref{energy}).  In general this time the winning point $\vr^*$
is not one-to-one with $\vw(\vr^*)$ even in the final convergent phase. 
Instead, the ``winner-takes-all" [UPDT:1] ensures one-to-one correspondence
relation between $\vr^*$ and $(\vw(\vr^*),w_{\cal O}(\vr^*))$.  Generally,
since our orientation data take only discrete values $\vxio={\rm e}^{{\rm i}\pi
k/N}\  (k=0,1,\dots,N-1)$, $\vwo(\vr^*)$ approaches locally constant functions
at the final convergent stage, i.e., $|\vxio -\vwo(\vr^*)|\rightarrow 0$ (where
local continuity of $\vwo(\vr)$ is assumed).  Thus the data-feature space
points  $(\vw(\vr^*),w_{\cal O}(\vr^*))$, in which the first coordinates
$\vw(\vr^*)$ is the same, are distinguished by the discrete ``labeling"
$\vwo(\vr^*)$.  Then our data profile ${\cal F}_{\vr^*}$  consists of a set of
pair $\{ (\vw(\vr^*),w_{\cal O}(\vr^*)) \}$ at the convergent phase.  Using
this fact, we can expand our energy function (\ref{energy}) at lowest order of
$\sigma_0$ as follows: 
\begin{equation}
\Delta{\cal E}\propto \lambda^{-1}\sigma^2_0\int_{\rm hypercolumn}
P(<\vw>)|J(\vv(\vr))|\left( (\partial\vv(\vr))^2 + g_1 (\partial
\vwo(\vr))^2\right){\rm d}^2\vr.
\label{hyperenergy}
\end{equation}
This gives an extra energy per hypercolumn.  Recall that $\lambda$ is a
measure for strength of the lateral correlation $\Lambda_{\vr,\vr^*}$. To avoid
confusion we write $\sigma_0\vv(\vr)\equiv\vw(\vr)\ - <\vw>$ instead of
$\vw(\vr)$ that was used for global retinotopic problem.  The global
retinotopic solution $<\vw>$ (equation (\ref{retino})) and the probability
$P(<\vw>)$ are treated almost constant within the hypercolumn region.   In the
above expression (\ref{hyperenergy}) we have to be careful about sign change
of the Jacobian $J(\vv)$.  This is because in comparison to the previous case
the relation between $\vr^*$ and $\vv(\vr^*)$ is not one-to-one; the Jacobian
could vanish and even change sign, giving rise to folding maps (see the
followings).

In order to get a picture for the ground state let us consider the possible
situations that lower the energy (\ref{hyperenergy}).  First of all, all
terms in the energy function are non-negative.  The lowest value in the
integrand is attained when the Jacobian factor $J(\vv(\vr))$ vanishes and the
locally constant function $\vwo(\vr)$ does not change its value
$|\partial\vwo|\neq\infty$ there.  In 2-dimensional mappings the vanishing
Jacobian $J(\vv(\vr))=0$ occurs in the following cases.
\begin{enumerate}
 \item[(i)] The case rank$[J]=1$
\begin{enumerate}\item[(i-a)] $J(\vv)$ changes sign in the neighborhood of 
                              $J(\vv)=0$.
                 \item[(i-b)]  $J(\vv)$ does not change sign in the 
                              neighborhood of $J(\vv)=0$.
\end{enumerate}
 \item[(ii)] The case rank$[J]=0$
\end{enumerate}
Here, abusing notations, we write  the Jacobian $J(\vv)=\det(
\partial v_i(\vr)/\partial r_j)\equiv\det [J]$.  The case (i) corresponds to
line singularities.  The equation of the singular line is determined by solving
$ J(\vv)=0$.  In detail, the subcase (i-a) indicates that there is a fold along
that line since changing the sign of the Jacobian implies alteration of
orientation  of the surface.  The subcase (i-b) means existence of a
sort of degeneracy
\bigskip\bigskip

\let\picnaturalsize=N
\def\picsize{5in}
\def\picfilename{figure2.eps}
\ifx\nopictures Y\else{\ifx\epsfloaded Y\else\relax \fi
\let\epsfloaded=Y
\centerline{\ifx\picnaturalsize N\epsfxsize \picsize\fi
\epsfbox{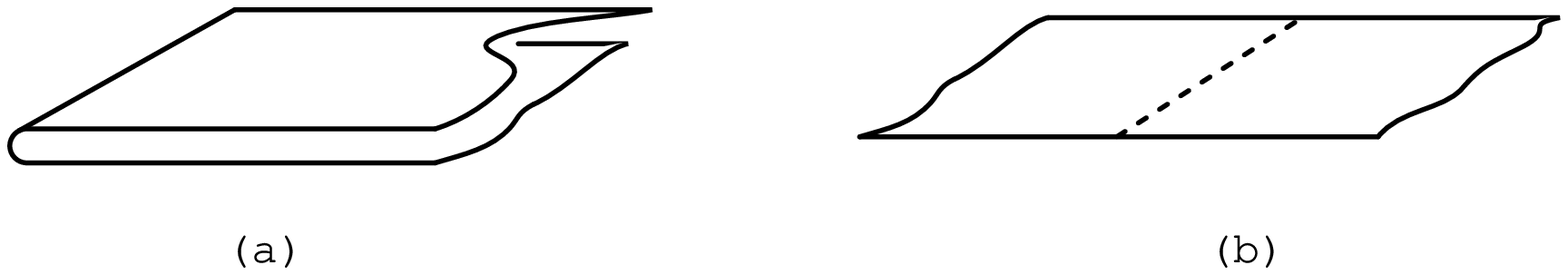}}}\fi

\noindent
of the mapping $\vv(\vr)$ along the line, but no folds.  The
simple example is shown in the Figure 2.   The sample mapping
$(v_1(\vr),v_2(\vr))=(r_1^2,r_2)$ in Figure 2$(a)$ exhibits a fold along the 
line $v_1=r_1=0$.  Figure 2$(b)$ shows a degenerate line singularities along
$v_1=r_1=0$ in the mapping $(v_1(\vr),v_2(\vr))=(r_1^3,r_2)$.

In our model (\ref{hyperenergy}) the type (i-a) singularity or a fold does not
occur unless the orientation function $\vwo(\vr)$ changes its (locally constant)
value when one crosses that folding line.  This is because the
``winner-takes-all" [UPDT:1] prohibits multiple winning points
$(\vw(\vr^*_i),\vwo(\vr^*_i))= (\vw(\vr^*_j),\vwo(\vr^*_j)),\
(\vr^*_i\neq\vr^*_j)$ that could otherwise occur
\smallskip
\let\picnaturalsize=N
\def\picsize{4.5in}
\def\picfilename{figure3.eps}
\ifx\nopictures Y\else{\ifx\epsfloaded Y\else\relax \fi
\let\epsfloaded=Y
\centerline{\ifx\picnaturalsize N\epsfxsize \picsize\fi
\epsfbox{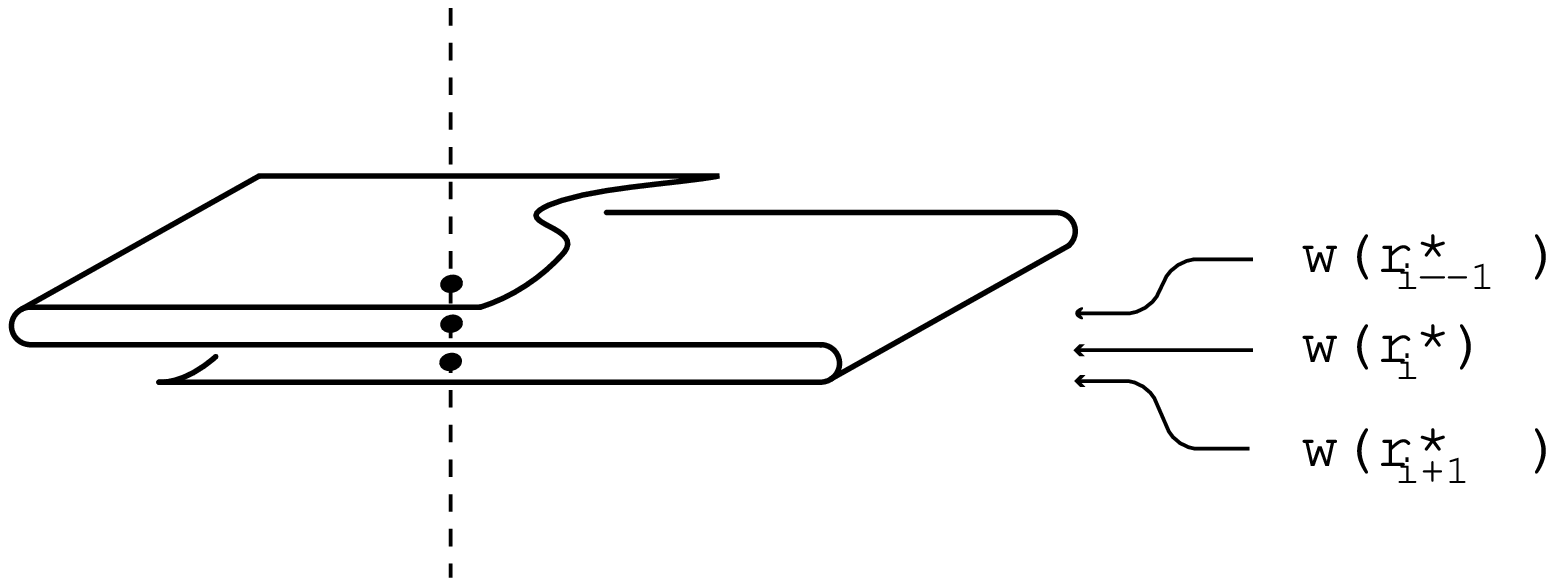}}}\fi
 
\medskip  
\noindent
on leaves $i,j$ of the folded
region. (See Figure 3.)  To prevent this at least each leaf has to be organized
to have different labels $\vwo(\vr^*_i)\neq\vwo(\vr^*_j)$ for $i\neq j$. Then
the contribution to our energy function (\ref{hyperenergy}) along these folding
lines is $(J=0)$ times $(|\partial\vwo|^2=\infty)$.  We encountered similar
situation in the previous work I.  We can employ a similar regularization method
in this case to make this product finite.  The lower energy configuration is
such that the total length of the folding lines is allowable shortest.  However,
this argument is based solely on the neighborhood of the folding lines. 
Actually the folding maps are unlikely to occur.  This is because, following the
analysis in the previous work I, locally at quite small scale the lower energy
mapping is almost identity mapping $\vw(\vr)\sim \vr$ up to conformal
rescaling.  Unless one feeds the system with non-homogeneous data in such a way
that a certain localized region gets mapped to be twisted with opposite
orientation, the folding is quite exceptional situation.  If it happens, the
folding line could be observed like ``fractures".

In contrast, the type (i-b) singularity is rather harmless.  It can occur
basically anywhere. In practice in computer simulations (or in real neuronal
circuits) this type of line singularity, which manifests itself as thin
accumulations of points onto a line in continuum theory, will not have 
observable effect, since the system is discretized.  Unless the lattice spacing
(or resolution) is quite small, the effect will be negligible.  If it produces
observable accumulation (or a certain area fails to grow and collapses to a line
by some reason), then this line would also resemble the ``fracture".

However, it if occurs at the boundary of iso-orientation region
($|\partial \vwo|=\infty$), it contributes to the energy function in the same
way as in the previous case (i-a).  In fact, this case is precisely the 
situation at the boundary of each ocular domains in Figure 2$(a)$ of I. There
we described the regularized value of the product $(J(\vw)=0)\cdot(|\partial
\vwo|=\infty)$ as kink energy.   The contribution of this to the energy
(\ref{hyperenergy}) gets lowered if the total length of the iso-orientation
boundary is  allowable, shortest.   

Here we should also point out that in the ground state the
iso-orientation boundary $|\partial \vwo|=\infty$ only occurs at those points
where the Jacobian vanishes $J(\vv)=0$. (Otherwise, the energy
(\ref{hyperenergy}) diverges and the corresponding probability distribution
vanishes.)  Therefore, the iso-orientation boundary typically does not have
unnecessarily wiggly shape, rather it is close to geodesic lines.

Finally, we have the case (ii) rank[$J$]=0.  In this case the condition
$J(\vv)=0$ determines a point, leading us to a point-singularity or an
intersection of line-singularities.  Before discussing the property of the
point-singularity, we briefly consider the property of non-singular region
$J(\vv)\neq0$.  There $\vwo(\vr)$ has to be almost constant
($\partial\vwo\approx0$) in the ground state.  Then the remaining first term in
energy (\ref{hyperenergy}) exhibits the same dynamics as the retinotopy problem
analyzed before.  Since the probability $P(<\vw>)$ is constant here, we can
reasonably assume so is $J(\vv)$, and $\vv(\vr)$ is locally harmonic
$\partial^2\vv=0$.  This could only be violated near the singular lines and
points where $J(\vv)$ approaches $0$.

Generally, using complex coordinates $z=r^1+{\rm i}r^2$, $\Phi=v^1+{\rm i}v^2$,
the harmonic function here can be written in a linear combination of holomorphic
and anti-holomorphic functions:
$$
 \Phi=\Phi_0(z)+\Phi_1(\bar{z}).
$$
Since at global scale the mapping is given in terms of an holomorphic (or
anti-holomorphic) functions (\ref{retino}), let us assume for the moment that
this property persists at hypercolumn scale.  Namely, we consider $\Phi(z)$ as
some sort of analytic continuation from global retinotopic function
(\ref{retino}).  This assumption is consistent with the mapping without folds.

In 2-dimensions the holomorphic mapping is (locally) a conformal mapping.  With
an appropriate coordinate transformation, the point singularities (of the type 
we are considering) can always be put into a standard form, e.g, near the
singular point at $z_0$ it reads
\begin{equation}
  \Phi(\vv(\vr))=(z-z_0)^{b(z_0)}.
\label{singular}
\end{equation}
Here the $b(z_0)\ (\geq 1)$ is called a
ramification index at that point.  (If $b(z_0)=1$, it represents a regular point
$J(\vv(z_0))\neq0$.)  This defines a branched  covering of the
retinal surface near the point $\vv(z_0)$; on the corresponding cortical surface
the $b(z_0)$ sheets of the covering  come together in
fan-shape.  By the same reason in the case of the folding maps (cf. Figure 3),
the ``winner-takes-all" ensures  a different labeling $\vwo(z)$ on each covering
sheet in the ground state.  (Therefore, the ramification index
$b(z_0)$ is smaller than possible orientation specifications $N$, i.e.
$b(z_0)\leq N$.)  Thus there should exist singular lines ($J(\vv)=0$) of type
(i-b) near this system.

The associated feature-space function $\vwo(z)$ is determined by minimizing the
energy (\ref{hyperenergy}) near that point.  Standard variation with respect to
$\vwo(z)$ gives the equation:
$$
\partial_z\overline{\partial}_z\vwo(z)+
J(\vv)^{-1}\partial_zJ(\vv)\cdot\overline{\partial}_z\vwo(z)+
J(\vv)^{-1}\overline{\partial}_zJ(\vv)\cdot\partial_z\vwo(z)=0.
$$
Substituting (\ref{singular}) into
$J(\vv)=|\partial_z\Phi(z)|^2-|\overline{\partial}_z\Phi(z)|^2$, we have
\begin{equation}
\partial_z\overline{\partial}_z\vwo(z)+
\frac{b-1}{z}\ \overline{\partial}_z\vwo(z)+
\frac{b-1}{\overline{z}}\ \partial_z\vwo(z)=0,
\label{w}
\end{equation}
where we put $z_0=0$, $b(z_0)=b$ for simplicity.  Since $\vwo(z)$ is the locally
constant function, this is always satisfied in almost all the places except near
singular points $J(\vv)=0$.  (See  Figure 4.)  As stated above each
covering-sheet has different label
$\vwo={\rm e}^{{\rm i}\pi k_i/N}$, $k_i\neq k_j$.  Since the
distance between the two points in the feature space

\let\picnaturalsize=N
\def\picsize{3in}
\def\picfilename{figure4.eps}
\ifx\nopictures Y\else{\ifx\epsfloaded Y\else\relax \fi
\let\epsfloaded=Y
\centerline{\ifx\picnaturalsize N\epsfxsize \picsize\fi
\epsfbox{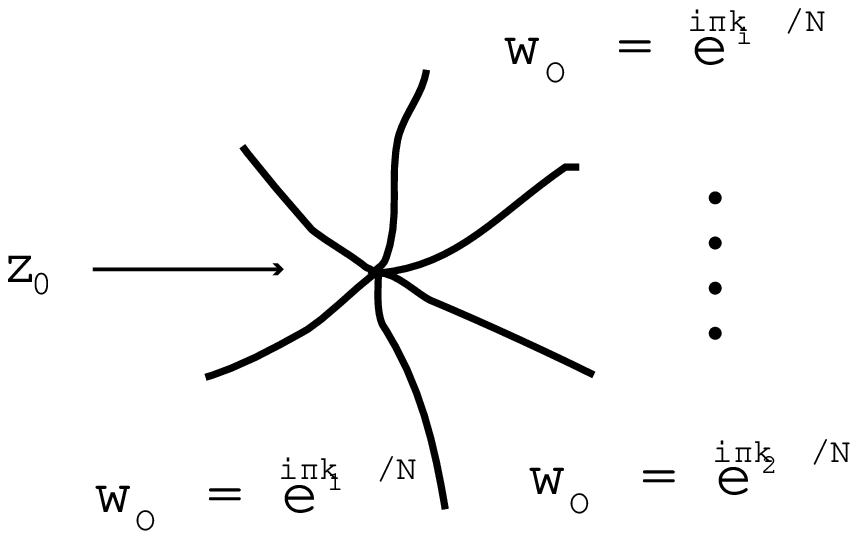}}}\fi

\medskip 
 
\noindent
is $|{\rm e}^{{\rm i}\pi
k_i/N} -  {\rm e}^{{\rm i}\pi k_j/N}|=2 \sin [\pi(k_i-k_j)/2N]$, the
neighbouring labels $k_i$ and $k_{i\pm1}$ is expected to take close values under
the [UPDT:2].   In the 
case $N\rightarrow\infty$
and the ramification index $b\rightarrow\infty\ ({\rm recall}\ N\geq b)$ where
the each covering sheet looks thin fan-shaped region on the cortical surface,
each $k_i$ approaches continuous function with respect to the index $i$, --- or
equivalently the  the function of $\arg[z]$.

In fact, one can solve equation (\ref{w}) generally as follows.  Since equation
(\ref{w}) can be rewritten as 
$$
  \partial_z\overline{\partial}_z
(z^{b-1}\bar{z}^{b-1}\vwo(z))=(b-1)^2(z\bar{z})^{-1}
(z^{b-1}\bar{z}^{b-1}\vwo(z))
$$
with $\partial_z\overline{\partial}_z$ being simply a 2-dimensional Laplacian, 
we can write down its general solution as a linear combination of products of
azimuthal functions ${\rm e}^{2{\rm i}\gamma\phi}=(z/\bar{z})^\gamma$ and radial
ones $f_\gamma(\rho)$ (where $\rho^2=z\bar{z}$) for various spectra $\gamma$. 
Explicitly we  have 
\begin{equation}
\vwo(z)=\int^\infty_{-\infty}{\rm d}\gamma\  A(\gamma) \left(\frac{z}{\bar{z}}
\right)^\gamma\!\cdot (z\bar{z})^{-(b-1)+\sqrt{\gamma^2+(b-1)^2}},
\label{expansion}
\end{equation}
where $A(\gamma)$ is an unknown  coefficient function (that would be
determined by solving global system including ocular terms in the energy
function).  In this equation we have discarded the other possible term
$(z/\bar{z})^\gamma (z\bar{z})^{-(b-1)-\sqrt{\gamma^2+(b-1)^2}}$, that is
divergent as $z\rightarrow0$, since $|\vwo(z)|\leq 1$ by definition.  Among the
expansion terms in equation (\ref{expansion}), only one term is dominant at
$z\approx0$; that is given by  
\begin{equation}
  \vwo(z)\ \approx\ A(\gamma_{\rm |min|})\left(\frac{z}{\bar{z}}
  \right)^{\gamma_{\rm |min|}^{}}\!\cdot (z\bar{z})^{\gamma^2_{\rm
  |min|}/2(b-1)}.
\label{wheel}  
\end{equation}
Here $\gamma_{\rm |min|}$ is the minimum positive or
maximum negative value of $\gamma$ (whichever the absolute value is smaller)
for which the expansion coefficient function $A(\gamma)$ does not vanish.  We
also took a leading term in $b\rightarrow\infty$ limit.  Depending on the
signature of $\gamma_{\rm |min|}$, $\vwo(z)$ gives clock-wise or anti-clock-wise
configuration of the orientation detector cells.  [If $A(0)\neq0$, then
$\gamma_{\rm |min|}=0$, i.e., the feature function $\vwo(z)$ is constant.
Following preceding argument below equation (\ref{w}), this is not the
case, however.]   Therefore, we conclude generally that in association to
singular points $\Phi(z)\sim z^b$ we have clock-wise or anti-clock-wise
orientation wheels, and that each orientation module organized around the
singular point is one of $b(z_0)$-tuple covering of the local area of the
retina.  Since in equation (\ref{wheel}) $\vwo(z)$ is a continuum limit
$N\rightarrow\infty$ of $\vwo={\rm e}^{{\rm i}\pi k_j/N}$, the $\gamma_{\rm
|min|}$ has to satisfy $-1/4\leq\gamma_{\rm |min|}\leq1/4$.

Now what so remarkable about working in (piecewise) holomorphic environment is
that there exist a celebrated topological formula by Riemann and Hurwitz that
relates the ramification indices $b(z)$ to the Euler characteristics of the
surfaces under consideration.  Specifically, let $\chi(E)$ and $\chi(F)$ be the
Euler characteristics of the surfaces $E$ and $F$, and $b(z)$ be the 
ramification index at $z\in F$ associated to the covering $\Phi(z):
F\rightarrow E$, then it states the relation
\begin{equation}
  \chi(F)=n\cdot\chi(E) -\sum_{z\in F}(b(z)-1).
\label{rh}
\end{equation}
In this equation $n$ is the mapping degree (or the number of the total coverings)
of the map $\Phi(z)$, i.e.,
$$
   n=\sum_{z\in \Phi^{-1}(w_0)} b(z)
$$
for an (arbitrary) point $w_0$ in $E$.  (For an accessible exposition on this
formula, see, e.g., Griffiths, P. and Harris, J., Principles of Algebraic
Geometry, John Wiley \& Sons, pp. 216-218 (1978).)  In the case at hand $E$ and
$F$ are small square regions of retina and visual cortex, respectively, and we
can take $\chi(E)=1$ and $\chi(F)\approx$ number of blocks of continuously
changing orientation area (see next section).  Plainly speaking, here the 
mapping degree is just the total number of orientation modules within one
hypercolumn.  Since the ramification index equals one at almost all the points
on the cortical surface $F$ (i.e., ordinary points), the summation $\sum_{z\in
F}(b(z)-1)$ in  equation (\ref{rh}) is actually a finite sum over singular
points to which orientation modules come together.  The most important thing
that equation (\ref{rh}) tells us is the ``existence" of singularities, namely,
since  the Euler characteristics are stated as above, as long as the number of
orientation module is larger than that of blocks of continuously changing
orientation area within one hypercolumn ($n> \chi(F)$), there must exist points
with non-trivial ramification index $b(z)>1$.  Following the preceding analysis
on the associated feature space function $\vwo(z)$, that would imply the
existence of the orientation wheels.

Unfortunately, global configuration of these singular points are beyond the
scope of this formula.  However, we can consider statistical distribution
function (a partition function) of possible mixed branched- and
unbranched-coverings of retinal surfaces, and write them down as a function of
number of singular points.  We will do this in the next section.

To summarize, we have discussed various possible situations that lower the
energy (\ref{hyperenergy}).  Semi-global structure at the scale of hypercolumn
size can be sketched as follows. Although there is a difference in
2-dimensional Gaussian factor $\Lambda_{\vr,\vr^*}$ here and the 1-dimensional
one used in the previous calculation in I, basic organization-structure in
arbitrary one-dimensional cross-section of the present case  will bear
resemblance to the kink-anti-kink configuration (Figure 2$(a)$ of I).  Namely,
dominance by only one feature (i.e., $\vwo(z)={\rm constant}$) over a broad 
range of the cortical surface is unlikely to occur.  This includes the case
$\vwo(z)=0$ where there is no orientation preference, since  these 
configurations merely raise the total energy by the same reason as in
1-dimensional case, i.e., the non-trivial contribution from the second term in
energy (\ref{energy}) gives large value proportional to the area of such
regions.  Instead the lower energy configuration is such that the various
locally constant values $\vwo(z)={\rm e}^{{\rm i}\pi k_j/N}$  are distributed 
over the surface in such a way that their boundaries (where
$|\partial\vwo|=\infty$) become (allowable) shortest and consistent with the
distributed orientation wheels.  The reason for this is that the boundaries
contribute finite positive energy after the regularization of the product with
Jacobian $(J(\vv)=0)\cdot(|\partial\vwo|^2=\infty)$, which is the same
situation as in I where the corresponding product was referred to as kink
energy.  

We also discussed the structure of point singularities in detail, and how
orientation wheels (both clock-wise and anti-clock-wise) occur in association
with the point singularities of the Jacobian $J(\vv)=0$.  The monotonic
(increasing or decreasing) change of the orientation preference angle was
also shown to take place as locally energy-minimizing configurations.  Most
remarkable is the number of singularities that was found to follow the 
celebrated Riemann-Hurwitz formula.   

\section{Partition function}

In this section we would like to estimate statistical distribution function of
our model associated to the lower energy configurations described in the
previous section.  This corresponds to evaluation of path integral (see, e.g.,
our previous work I)
\begin{eqnarray}
{\cal Z}&\sim& \int {\cal D}\bv\ \ {\rm e}^{-\Delta {\cal  E}
[\mbox{\boldmath\scriptsize $v$}]}\nonumber \\
  &\sim & \sum_{\mbox{\boldmath\scriptsize $v$}} \left[ {\rm multiplicity\  of\
the\ covering\ map}\ \bv \right]\ {\rm e}^{-\Delta {\cal 
E}[\mbox{\boldmath\scriptsize $v$}]} \nonumber 
\end{eqnarray}
for the dominant contributions.  As we discussed before the covering map here
is specified by its mapping degree $n=N$, the total number of branch points $I$,
their positions $z_i$, and associated ramification indices $b(z_i)$.  Most of
this section is devoted to the counting problem of the multiplicities of these
mappings for the given data $(N,I,{b(z_i)})$.

\subsection{Unbranched and branched coverings of retinal surface}

Although in this paper our main concern is the covering maps of an area
of retina, that is topologically equivalent to a disk, we first start with an
unbranched covering of a torus for its illustrative purpose.  We write a
plaquette for a torus, assuming identification (Figure 5$(a)$) of
like-labeled edges in each side.  An unbranched $N$-sheeted covering is a
collection of $N$ such objects glued together (Figure 5$(b)$) along
like-labeled edges.  There are many varieties to construct this type of
covering surfaces (including  ones with disconnected
components).  In order to cover the original torus, one identifies each
horizontal edges, and then wraps up the torus with the
resulting 
\vspace{10pt}
\let\picnaturalsize=N
\def\picsize{4.5in}
\def\picfilename{figure5a.eps}
\ifx\nopictures Y\else{\ifx\epsfloaded Y\else\relax \fi
\let\epsfloaded=Y
\centerline{\ifx\picnaturalsize N\epsfxsize \picsize\fi
\epsfbox{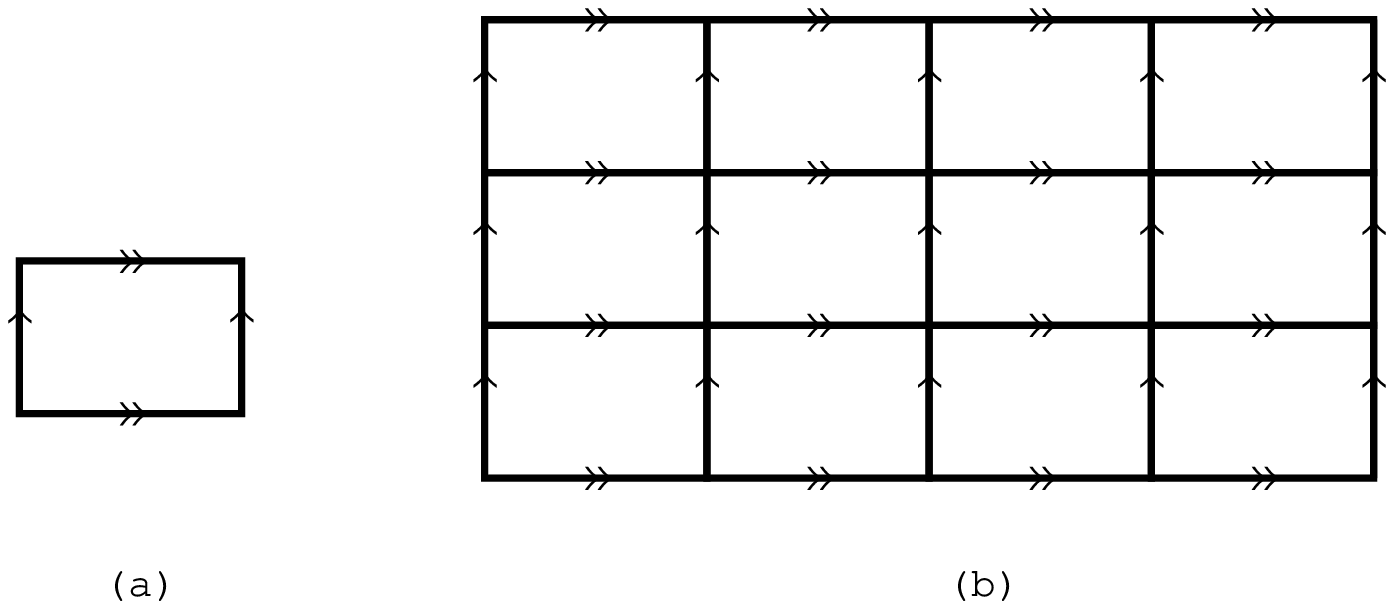}}}\fi

\medskip 
 
\let\picnaturalsize=N
\def\picsize{2in}
\def\picfilename{figure5c.eps}
\ifx\nopictures Y\else{\ifx\epsfloaded Y\else\relax \fi
\let\epsfloaded=Y
\centerline{\ifx\picnaturalsize N\epsfxsize \picsize\fi
\epsfbox{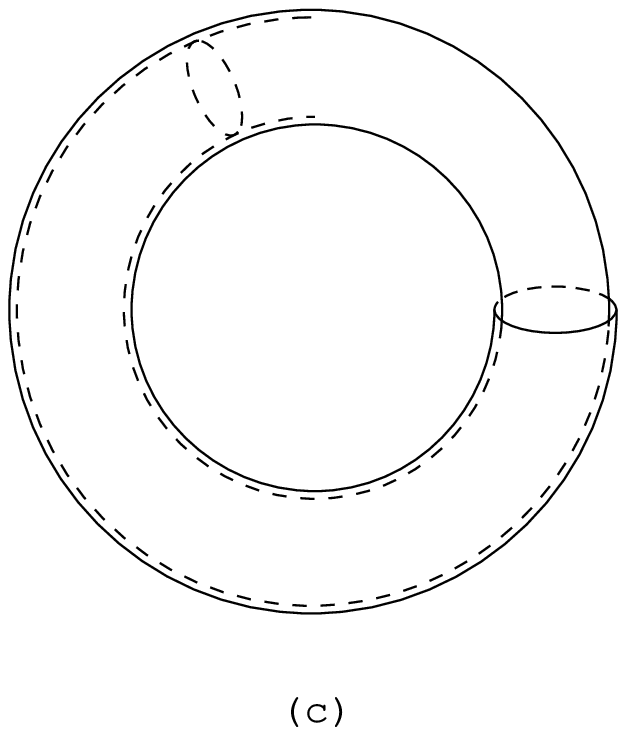}}}\fi
\vspace{-5pt} 
\begin{center}
{\footnotesize {\bf Figure 5.}  
(cont'd).} 
\end{center}
\medskip 

\noindent
 cylinder by putting one end of the cylinder in the other end until
the circumferences match up with the original torus (Figure 5$(c)$).  In this
way, every point of the original torus are covered precisely $N$ times.  The
Riemann-Hurwitz formula (\ref{rh}) is trivially satisfied.

Next we consider a branched covering of a sphere ($S^2$).  We start with a
2-sheeted unbranched covering of $S^2$.  Just as in previous case the
unbranched covering corresponds to a simple double wrapping of $S^2$; every
points are covered twice in this covering.  We name the covering sheets ``1"
and ``2".  To make from this a branched covering, we pick two points $(A,B)$ on
the sphere and cut the covering sheets along the straight line connecting these
two points (Figure 6$(a)$).  Then we identify one edge of the sheet ``1" with
the edge of the sheet ``2" on the other side of the cut and the remaining edge
of the sheet ``1" with that of the sheet ``2", i.e., identification of the
like-labeled edges $p,q$ in Figure 6$(a)$.   The resulting covering is
branched; the two points $A,B$ we picked are branch points (of index 2). 

To see its
topologically equivalent surface, we peel off the outer sphere, and paste it
back along the edges $p,q$ as prescribed (Figure 6$(b)$).  Then we end up with 
an object in  sphere topology.  The Riemann-Hurwitz formula (\ref{rh}) is
satisfied since the Euler indices $\chi(E)=\chi(F)=2$ for the sphere, $n=2$
(double covering), and the ramification indices $b=2$ at both branch points
$(A,B)$.  [In the case of simple unbranched covering the Euler characteristic
of the covering surface is just the multiple of the original one by the number
of coverings.] 
  
\bigskip
\let\picnaturalsize=N
\def\picsize{5in}
\def\picfilename{figure6.eps}
\ifx\nopictures Y\else{\ifx\epsfloaded Y\else\relax \fi
\let\epsfloaded=Y
\centerline{\ifx\picnaturalsize N\epsfxsize \picsize\fi
\epsfbox{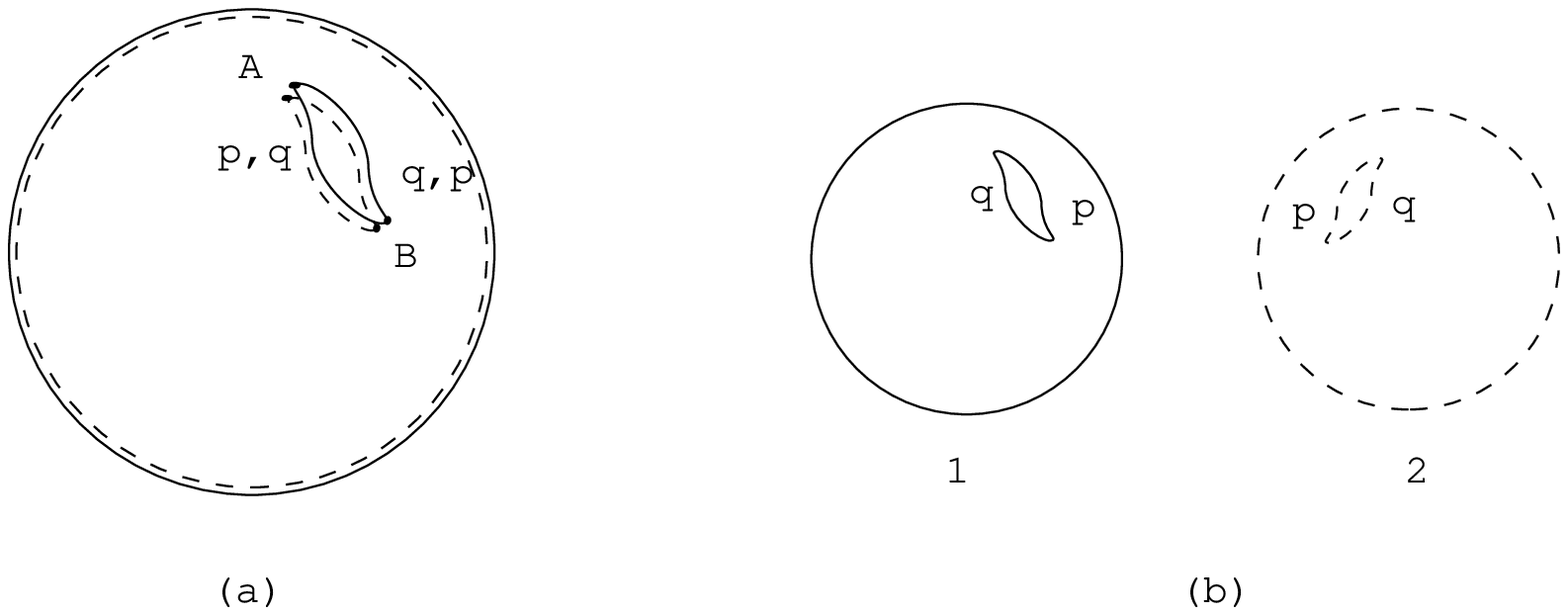}}}\fi

\medskip

An example of the corresponding analytic expression for this covering is the
mapping $z\mapsto w\equiv z^2(\in {\Bbb C}\cup\{\infty\})$, where the
singularities (of the ramification index 2) occur at the north ($z=\infty$) and
south ($z=0$) poles of the (stereo-graphically projected) sphere.

The branch points on this sphere are characterized as follows.  We draw a
small circle on the sphere surrounding one of the branch point.  Obviously,
going around once on the circle takes one from the covering-sheet ``1" (or ``2")
to ``2"(or ``1"), and a repetition of this process brings one back to the
starting position. The same is true with regard to the other branch point.  One
can think of this process as an action of transpositional element (12) in
the symmetric group ${\cal S}_2$ on the labels ``1, 2" of the covering-sheets.

For the case 3-sheeted branched covering with two branch points, we can repeat
the similar construction, using identification rules as shown in Figure 7$(a)$.
The Figure shows the sectional view from the north pole.  It is easy to see
that the resulting branched cover has again sphere topology.  An example of the
corresponding analytic mapping is $z\mapsto w\equiv z^3$ with the
ramification indices $b(0)=b(\infty)=3$.  In this case, the branch points are
characterized by the permutational element (123) of symmetric group
${\cal S}_3$.  Figure 7$(b)$ shows another gluing rule in which mixed branched
and unbranched coverings occur.  This time the branch points are
characterized by the transposition (12)(3) of the ${\cal S}_3$.  In Figure 
7$(c)$ in another construction we separated out the two branch points with
ramification indices 3 in Figure 7$(a)$ into four branch points with
ramification indices 2.  [The total branching number is invariant in this
process: $(3-1)\times 2 =(2-1) \times 4$.] \ \ The new branch points are
characterized by (12)(3) and (1)(23) of the symmetric group ${\cal S}_3$,
respectively, associated to each gluing edge.

\let\picnaturalsize=N
\def\picsize{4.5in}
\def\picfilename{figure7.eps}
\ifx\nopictures Y\else{\ifx\epsfloaded Y\else\relax \fi
\let\epsfloaded=Y
\centerline{\ifx\picnaturalsize N\epsfxsize \picsize\fi
\epsfbox{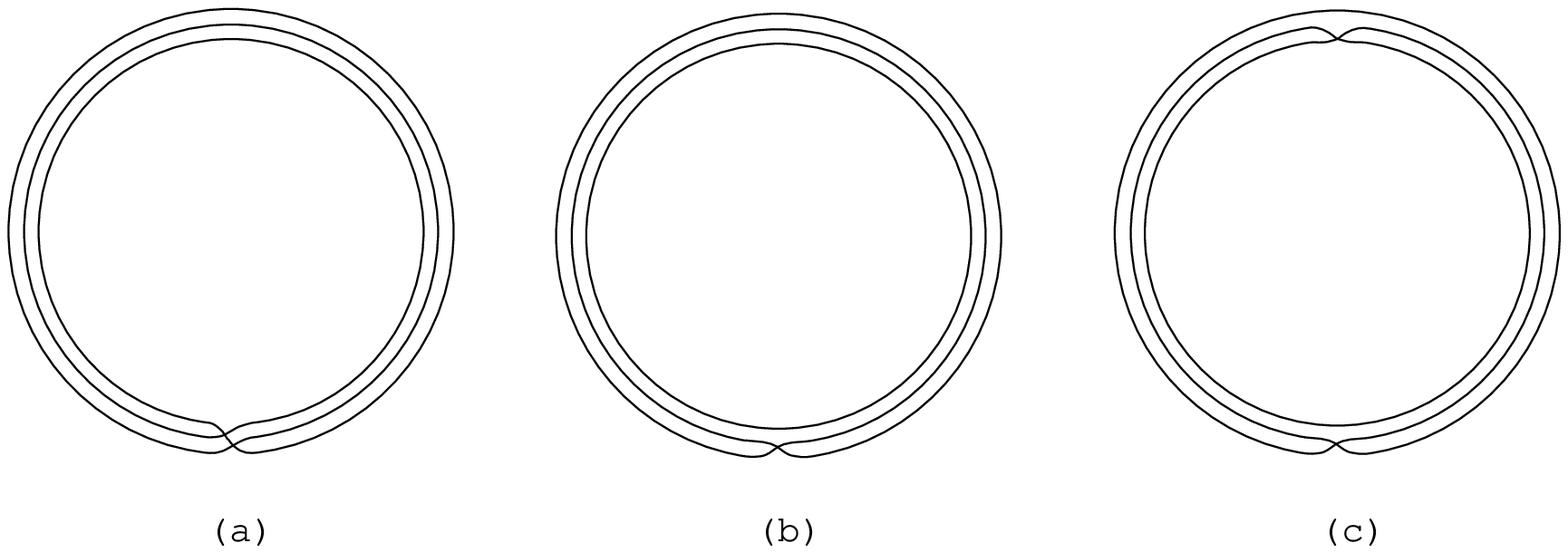}}}\fi
 \medskip

One can generalize this construction, and create $N$-sheeted branched (and
unbranched) coverings with $I$ branch points by associating permutation
elements $\pi_i,\ i=1,\dots,I$ in symmetric group ${\cal S}_N$ to each branch
points $z_i$ on the sphere.  The choice of the permutations $\pi_i$'s is
basically arbitrary except one constraint to be satisfied by $\pi_i$'s: 
\begin{equation}
   \pi_1\pi_2\cdots\pi_I=1.
\label{constraint}
\end{equation}
The reason for this is the following (e.g., Hurwitz 1891).  We draw small
circles $c_i$ around every branch points $z_i$, and go around them once starting
from an arbitrary non-branch point $P$ on the sphere (Figure 8$(a)$).  This
gives the left-hand side of the constraint equation (\ref{constraint}).  Since
we are on the sphere, we can deform this loop (without crossing any branch
points) into another loop with the opposite orientation that surrounds small
region near the starting point $P$ (Figure 8$(b)$).  Since the point $P$ is a
regular point, this gives the right-hand side of the equation
(\ref{constraint}).  One obtains different coverings according to the various
different combinations of $\pi_i$'s.
    
The number of independent coverings created in this way can be counted as
$$
   \sum_{\pi_i\in {\cal S}_N}{}^{\!\!\!\!\! '}\ \delta (\pi_1\pi_2\cdots\pi_I)
$$ 
where $\sum{}^{'}$ stands for the sum avoiding double counting, and
$\delta(p)$ is Kronecker's delta
$$
\delta(p)=\left\{ \begin{array}{ll}
                       1  & \mbox{if $p=1$} \\
                       0  & \mbox{otherwise} .
                  \end{array} \right.
$$
We will specify the counting rule $\sum{}^{'}$ in our
specific problem later.
  
\let\picnaturalsize=N
\def\picsize{5.1in}
\def\picfilename{figure8.eps}
\ifx\nopictures Y\else{\ifx\epsfloaded Y\else\relax \fi
\let\epsfloaded=Y
\centerline{\ifx\picnaturalsize N\epsfxsize \picsize\fi
\epsfbox{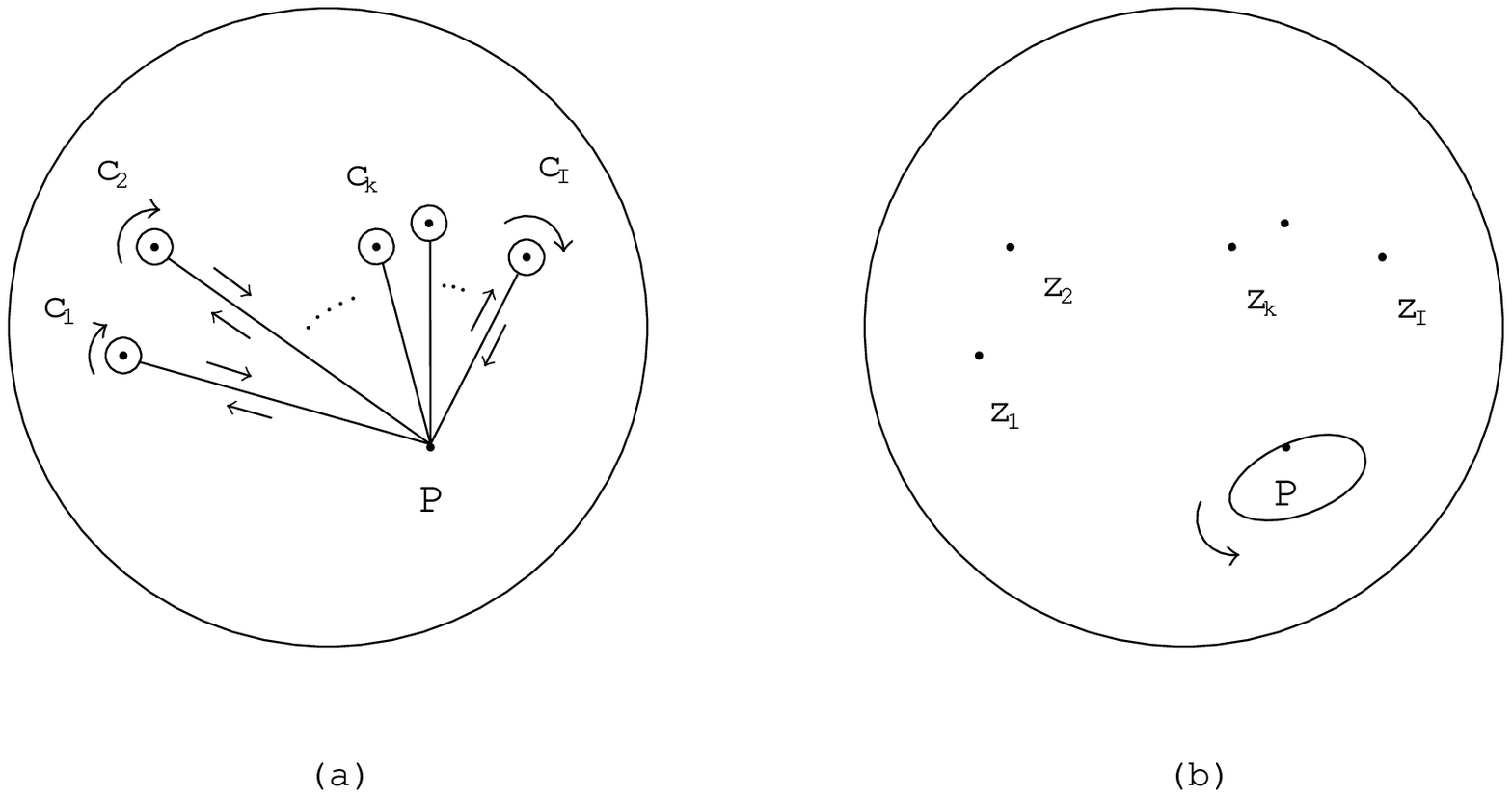}}}\fi
 
\medskip

Now  we would like to consider a similar construction for coverings of a
plaquette that is needed in this paper. In this case we have to take into 
account the coverings of the boundary.  We assume that our covering sheet(s)
have $m(\leq N)$ non-intersecting, mutually independent boundaries $B_k\
(k=1,\dots,m)$, each homeomorphic to a circle $S^1$, and that the boundary of
the original plaquette is covered by these boundaries $B_k$, each $L_k$ times,
for some non-negative integer $L_k$, ($L_1+L_2+\cdots+L_m=N$).  [Namely, $B_k$
is the $L_k$-fold cover of the boundary of the original plaquette.]

In the similar way as we specified branch points we can specify this
boundary covering in terms of an element $\sigma\in {\cal S}_N$ that consists of $m$
cycles of lengths $L_1,L_2,\dots,L_m$, i.e.,
$\sigma=(\ell_1\ell_2\cdots\ell_{L_1})(m_1m_2\cdots
m_{L_2})(\ \cdots\ )(q_1q_2\cdots q_{L_m})$.  If one goes around the boundary of
the original plaquette once in anti-clock-wise direction, one moves on the
boundary circles in covering space in a way as specified by $\sigma$.

With these preparations we can now construct a branched covering of a
plaquette in exactly the same way as in the sphere case.  However, here the
constraint (\ref{constraint}) is replaced by the following one:
\begin{equation}
   \pi_1\pi_2\cdots\pi_I=\sigma^{-1}.
\label{constraint2}
\end{equation}
The reason is obvious by construction; we just repeat the discussion following
equation (\ref{constraint}) with the sphere replaced by a plaquette.  

We get
various mixed branched (and unbranched) coverings under different combinations
of $\pi_i$'s and $\sigma$ in the symmetric group ${\cal S}_N$.  

In order for this construction to be applicable to our specific
problem of orientation modules, we must restrict possible forms of $\pi_i$'s and
$\sigma$.  Recall that every orientation modules are labeled by the locally
constant function $\vwo={\rm exp}({\rm i}\pi k_i/N),\ k_i=1,\dots,N$, and they
constitute an $N$-sheeted covering of an area of retinal surface.  Since,
as discussed in the previous section, the neighbouring modules have similar
orientation specifications $k_i$, first we must have the boundary $\sigma$
as in  
\begin{equation}
   \sigma=(12\cdots L_1)^{\epsilon_1}(L_1+1,\cdots,
L_1+L_2)^{\epsilon_2}(\quad\cdots\quad)(N-L_m+1, \cdots, N)^{\epsilon_m},
\label{sigma}
\end{equation}
where $\epsilon_i=\pm1\ (i=1,2,\dots,m)$.
Secondly, in order to prevent double counting we restrict all $\pi_i$ to be
transpositions $(k\ell)\in {\cal S}_N$.  This is because as we saw in Figure 7
non-transpositional gluing rule can always be considered as a special case
of  transpositional gluings where some of the branch points of the
ramification indices 2 accidentally come close by.  Furthermore in this case we
must restrict them to the generators of the symmetric group $(k\ k+1),\
k=1,2,\dots, N-1$, and $(N\ 1)$.  The reason for this is locally the
neighbouring modules have similar orientation specifications $k_{i\pm
1}=k_i\pm 1$.  

Putting all these together, we end up with the expression for the partition
function of our orientation modules of module number $N$ per
hypercolumn: 
\begin{equation}
  {\cal Z}_N\sim \sum^{\infty}_{I=0}\,\sum_{\sigma,\pi_i\in
{\cal S}_N}{}^{\!\!\!\!\!\!\!  '} \ \ 
\delta(\pi_1\cdots\pi_I\sigma)\int\prod^{I}_{i=1}{\rm
d}^2z_i\cdot |\Delta[\bv ]|\,{\rm e}^{-\Delta {\cal
E}[\mbox{\boldmath\scriptsize $v$}]}. \label{partition}
\end{equation}
Here the summation $\sum^{ '}_{\sigma,\pi_i\in {\cal S}_N}$ implies that we 
restrict $\pi_i$  to generators $(k\ k+1),\ k=1,2,\cdots,N-1$, and $(N\ 1)$
only, and $\sigma$ as in equation (\ref{sigma}) while avoiding any double
counting.  The energy $\Delta{\cal E}[\bv]$ is a function of singular points
$z_i$ and the boundary condition specified by $\sigma$.\footnote[3]{The 
quantity $\Delta[\bv]$ is a Jacobian arising from change of functional
integration variable ${\cal D}\bv\rightarrow{\rm d}^2z_i$ in collective
coordinate method.}

\subsection{Most probable modules}  
Although we are interested in the case where the total module number $N$ (per
hypercolumn) is large, let us first consider, as an example, the case $N=5$. 
First of all, since the cortical surface $F$ has only simple topology, as a
covering space its Euler characteristic $\chi(F)$ has to be a positive integer
no larger than $N$; $\chi(F)=1,2,3,\dots, N$.  The case $\chi(F)=N$
corresponds to $N$ unbranched (disconnected) coverings, and $\chi(F)<N$ if the
covering is of mixed branched and unbranched one.  From the Riemann-Hurwitz
formula (\ref{rh}) this restricts the number of (transpositional) singularities
$I=0,1,2,\dots,N-1$.  In the case $N=5$ this implies $I=0,1,\dots,4$, and we
can easily evaluate possible number of mappings
$\sum^{'}\delta(\pi_1\cdots\pi_I\,\sigma)$ (Table 1).    For the boundary
condition $\sigma=1$ we get $\sum^{'}\delta(\pi_1\pi_2)=5$,
$\sum^{'}\delta(\pi_1\pi_2\pi_3\pi_4)=15$, and etc.  Intuitively these surfaces
correspond to coverings connected with double arcs (see Figure 9$(b)$).  In
Figure $9(a)$ we start with $(N=)\,5$ unbranched covering-sheets 

\bigskip
\let\picnaturalsize=N
\def\picsize{5.5in}
\def\picfilename{figure9.eps}
\ifx\nopictures Y\else{\ifx\epsfloaded Y\else\relax \fi
\let\epsfloaded=Y
\centerline{\ifx\picnaturalsize N\epsfxsize \picsize\fi
\epsfbox{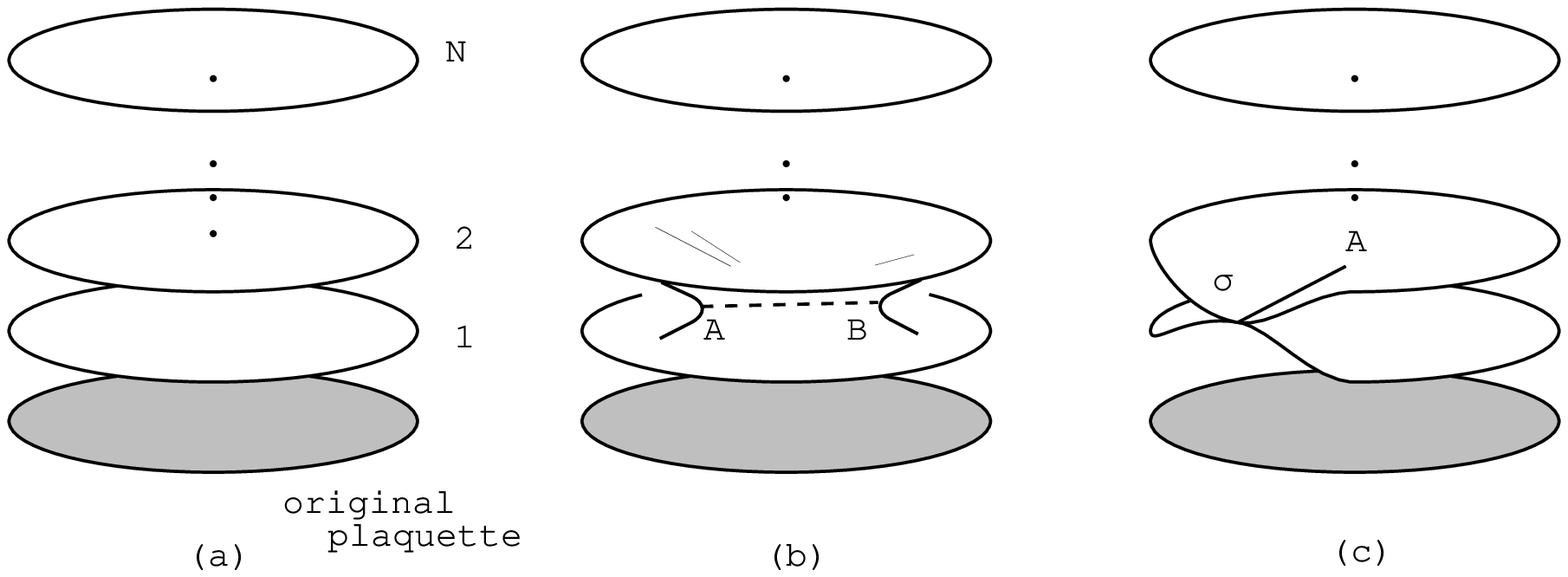}}}\fi
  \medskip 

\noindent 
of
the original plaquette.  In Figure $9(b)$, we connected the sheets ``1" and
``2" along a line connecting two singularities $A,B$, via $\pi_A=\pi_B=(12)$,
using a similar rule as explained in Figure 6 in the sphere case.  The line
connecting the two surface is called a ``double arc".  Every term of the type
$(i,i+1)\cdots (i,i+1)$ appearing in $\pi_1\pi_2\cdots\pi_I$ represents such a
double arc connecting a pair of covering-sheets ``$i$" and ``$i+1$".

For the case the boundary condition $\sigma=\sigma(j)\equiv(j,j+1)$, we get
$\sum_j\sum^{'}\delta(\pi_1\,\sigma(j))=5$,
$\sum_j\sum^{'}\delta(\pi_1\pi_2\pi_3\,\sigma(j))=15$, and etc.  In this case
the intuitive picture is just like the double arcs stated above except for one
singular point that is anchored to  $\sigma=(j,j+1)$ on the boundary (Figure
$9(c)$).  The latter comprises a branch cut along the line connecting them.  
The pairs of transpositions that do not contract with $\sigma$ compose the
double arcs as before.

In general we can locate $\sigma$ anywhere at the boundary.   As in the case of
singularities $\pi_j$ we consider the component-decomposition of $\sigma$ in
terms of transpositional elements $(j,j+1)$.  We assume that
these components  of  $\sigma$ are distributed separately over the
boundary in the same order appearing in the transpositional component
decomposition in minimal length.  The summation
$\sum^{'}_{\sigma\in{\cal S}_N}$ thus gives rise to  factors of $\zeta$(= length
of the boundary in the unit of $\sigma_0$(=\,hypercolumn size)) in association
with each transpositional component $(j,j+1)$.

\smallskip  
\begin{center}
\begin{tabular}{l|rrrrr} 
$\qquad\qquad\qquad I$ & 0 & {1\ } & {2\ \ \ } & {3\ \ \ } & {4\ \ \  } \\ \hline
$\sum^{'}\delta(\pi_1\cdots\pi_I)$ & 1 & & $5{\cal A}^2$ & & $15{\cal A}^4$ \\
$\sum_i\sum^{'}\delta(\pi_1\cdots\pi_I\,(i,i+1))$ &  & $5g$ &  & $15g {\cal
A}^2$ & \\ 
$\sum_i\sum^{'}\delta(\pi_1\cdots\pi_I\,(i,i+1,i+2))$ &  &  & $10g^2 $
&  & $20g^2 {\cal A}^2$ \\
$\sum_i\sum^{'}\delta(\pi_1\cdots\pi_I\,(i,i+1,i+2,i+3))$ &  &  &  & $10g^3 $
& \\    
$\sum_{\rm cycles}\sum^{'}\delta(\pi_1\cdots\pi_I\,(12345))$ &  & &  & &
$10g^4$ \\ 
$\sum_{\rm pairs}\sum^{'}\delta(\pi_1\cdots\pi_I\,(12)(34))$ & &
& $5g^2 $ & &  $10g^2 {\cal A}^2$  \\
$\sum_{\rm pairs}\sum^{'}\delta(\pi_1\cdots\pi_I\,(123)(45))$ &  &  &  & $10g^3$
& \\   
\end{tabular} 
\end{center} 
\bigskip

In the path integral (\ref{partition})
the integration $\int{\prod^{I}_{i=1}{\rm d}^2z_i}$ also gives rise to factors
of ${\cal A}_0$(= area of the covered plaquette in the unit of $\sigma_0^2$) in
association with the positional degrees of freedom of the (transpositional)
singularities.  [For the complete handling of these factors (proportional
to the boundary length and area), one actually has to solve stationary point
equation associated to (\ref{energy}), and to integrate over all the positions 
of the singularities and the boundary conditions; the integrand $e^{-\Delta
{\cal E}}$ depends on those (so-called moduli) parameters in general.  The
actual integration over the moduli  generally gives slightly different
contributions rather than the simple multiplications of factors of the area and
the boundary length of the plaquette as above.  However, here in order to get
rough estimation of the path integral we ignore those parameter dependence of 
the integrand $ e^{-\Delta {\cal E}}$ ($\Delta{\cal E}\equiv\Delta{\cal
E}_0/\lambda$, cf. equation (\ref{hyperenergy})), and we simply multiply the
factors of ${\cal A}_0$ and $\zeta$ to its representative value $e^{-\Delta
{\cal E}_0/\lambda}$ per (transpositional) singularity $(i,i+1)$.]  We combine
two factors ${\cal A}_0$ and $e^{-\Delta{\cal E}_0/\lambda}$, and write ${\cal
A}={\cal A}_0e^{-\Delta{\cal E}_0/\lambda}$. In Table 1 instead of the
variables $\zeta, {\cal A}$ we have used $g(\equiv\zeta{\cal A}), {\cal A}$.  
In the latter parametrization the power of $g$
equals (minimal) transpositional component number of the boundary
condition $\sigma$, and that of ${\cal A}$ represents twice the number of
double arcs in the resulting coverings.

From the Table 1 and the definition of ${\cal A}={\cal A}_0e^{-\Delta{\cal
E}_0/\lambda}$ one can immediately conclude that appropriate choice of the
lateral correlation strength $\lambda$ in our model suppresses (i.e., ${\cal
A}<1$) double-arc contributions.   We assume such choice of $\lambda$ in the
following, since that is close to real situation.  Then among the resulting
${\cal A}$-independent terms the maximum contribution is obtained from the term
with the boundary condition $\sigma=(12345)$-type if $g>1$, and from other
$\sigma$'s if $g<1$.  Since ${\cal A}_0$ is of order $1$ and
$\zeta\sim24\sqrt{3}$,\footnote[2]{The ocular-dominance width
$\sim4\sqrt{3}\sigma_0$ from our previous work (I) has been used.}  (unless the
measure $\lambda$ for the lateral correlation strength is  too small) the former
boundary condition $\sigma=(12345)$-type is realized.  Hence one gets one
orientation-wheel with all orientation directions organized to it.  (We assume
\underline{in the global scale} that every transpositional components (separated
as a convention for the counting multiplicities) are organized in such a way
that each boundary condition $\sigma$ (with $m$ cycles) corresponds to  global
$m$ orientation-wheel(s).)

The next largest contribution is from those terms with $\sigma=(1234)$-type or
$\sigma=(123)(45)$-type, namely, singularities compose either one wheel with
fewer orientation-modules or two orientation wheels with three- and
two-orientation specifications as in $(123)$ and $(45)$.  (When $g$ becomes
smaller than 1, the boundary conditions other than $\sigma=(12345)$-type give
dominant contributions.  For the detailed classification of this case we would
need more precise evaluation of each term in  Table 1.)

\begin{center}
\begin{tabular}{l|cccccccc} 
$\qquad\qquad\qquad I$ & 0 & {1\ } & {2\ \ \ } & {3\ \ \ }  &
{$\cdots$}& $L$ & $\cdots$ & {$N$}\\ \hline 
$\sigma=1$ & 1 & &  & & & & & \\
$\sigma=(i,i+1)$ &  & $Ng$ &  &  &  & & & \\ 
$\sigma=(i,i+1,i+2)$ &  &  & $2Ng^2 $ &  & & & &  \\
$\sigma=(i,i+1,i+2,i+3)$ &  &  &  & $2Ng^3 $ & & & & \\    
$\qquad \vdots$ &  & &  & & &$\cdots$ & & \\ 
$\sigma=(123\cdots N)$ & & &  & & & & & $\ 2Ng^{N-1}$  \\
$\sigma=(12)(34)$ & & & $N^2g^2/2$ &  & & & & \\ 
$\sigma=(12)(34)(56)$ & & &  & $N^3g^3/3!$ & & & & \\
$\qquad \vdots $ & & &   & & $\ddots$ & & &\\
$\sigma=\{ \ell_1,\ell_2,\dots,\ell_m;n\}$ & & & & & &
$Q_\sigma$ & &\\
\end{tabular}
\end{center}  
\medskip

Now we turn to the case $N$ large (Table 2)\footnote[3]{In the table we
write only representative boundary conditions $\sigma$ in each row; the
summation with respect to $i$ or possible $\sigma$'s that fall into the same
class as the representative $\sigma$ is assumed.}.  Basic structure of the
combinatorial factors remains same as in $N=5$ case.  Since we are not
interested in the coverings with multiple double-arcs, we omit the terms with
explicit ${\cal A}$ dependence in Table 2.  For this purpose we have assumed
the appropriate choice of the strength measure $\lambda$ of the lateral
correlation $\Lambda_{\vr,\vr^*}$ (so that ${\cal A}\equiv{\cal A}_0
e^{-\Delta{\cal E}_0/\lambda}<1)$.  We also
write the boundary condition $\sigma$ as $\{ \ell_1,\ell_2,\dots,\ell_m;n\}$ 
that consists of cycles of lengths $\ell_1,\ell_2,\dots,\ell_m$ among which $n$ 
cycles are non-transpositional.  Approximate combinatorial
factor $Q_\sigma$ for the general $\sigma=\{ \ell_1,\ell_2,\dots,\ell_m;n\}$ is
given as   
\begin{equation}
   Q_\sigma\approx 2^nC_\sigma Ng^L\prod^{m-1}_{k=1}\left(1+ \frac{N-L-m+1}{k}
\right)\, ,\label{combinatorial}
\end{equation}
where $C_\sigma$ is a symmetry factor, e.g., $C_\sigma=1/m$ if all the
cycle-length $\ell_i$ are equal, $C_\sigma=1/3$ if $\sigma=\{\ell_1,\ell_2,
\ell_1,\ell_2,\ell_1,\ell_2;n\}$, etc., and $C_\sigma=1$ if there is no
symmetry in the configurations of $\ell_i$'s, and
$L\equiv\sum_{i=1}^{m}(\ell_i-1)$.

A larger combinatorial factor is obtained if $n$ is large ($=m$) and $\sigma$ is
non-symmetric (so that $C_\sigma=1$).  Restricting $\sigma$ to such cases, we
evaluate the factor $Q_\sigma$ in the decreasing order of the power $g^L$.  In
general there are several configurations $\sigma$ that have  the same power
$g^L$.  For instance, in the case $L=N-4$ we have $\sigma=\{\ell_1,\ell_2,
\ell_3,N-\sum_{i=1}^{3}\ell_i\,;4\}$, $\{\ell_1,\ell_2,
N-1-\sum_{i=1}^{2}\ell_i\,;3\}$,   $\{\ell_1,N-2-\ell_1\,;2\}$, and $\{N-3;1\}$
(, assuming  $\ell_1\neq \ell_2\neq \ell_3 >2$, etc.).  The first $\sigma$
represents four orientation-wheels with all orientation modules belong to one
of the four wheels, whereas the remaining $\sigma$'s represent three, two, or
one orientation-wheel(s) with one, two, or three orientation modules
disintegrated from the organized wheel(s). The corresponding combinatorial
factors are given as $Q_\sigma=64Ng^{N-4}$, $48Ng^{N-4}$, $16Ng^{N-4}$, and
$2Ng^{N-4}$, respectively.  Thus the maximal factor $Q_\sigma$ is attained in
the first $\sigma$.  In Table 3 we list only those maximal
configurations $\sigma$ together with their combinatorial factors $Q_\sigma$
for each $g^L$.

\begin{center} 
\begin{tabular}{l|r} 
$\qquad\qquad \sigma$ & {\hbox to1in{\hfil$Q_\sigma$\hfil}}\\ 
\hline 
$\{N;1\}$ & $2Ng^{N-1}$ \\
$\{\ell_1,N-\ell_1;2\}$ & $8Ng^{N-2}$ \\
$\{\ell_1,\ell_2,N-\sum_{i=1}^{2}\ell_i\,;3\}$ & $24Ng^{N-3}$ \\
$\{\ell_1,\ell_2,\ell_3,N-\sum_{i=1}^{3}\ell_i\,;4\}$ & $64Ng^{N-4}$ \\ 
\hline
$\{\ell_1,\ell_2,\ell_3,N-1-\sum_{i=1}^{3}\ell_i\,;4\}$ & {\vbox
to0pt{\vskip0pt\hbox to1in{\hfil$160Ng^{N-5}$}\vfil}}\\
$\{\ell_1,\ell_2,\dots,\ell_4,N-\sum_{i=1}^{4}\ell_i\,;5\}$ &  \\
\hline $\{\ell_1,\ell_2,\dots,\ell_4,N-1-\sum_{i=1}^{4}\ell_i\,;5\}$ &
$480Ng^{N-6}$ \\ $\{\ell_1,\ell_2,\dots,\ell_5,N-1-\sum_{i=1}^{5}\ell_i\,;6\}$
& $1344Ng^{N-7}$ \\ \hline
$\{\ell_1,\ell_2,\dots,\ell_5,N-2-\sum_{i=1}^{5}\ell_i\,;6\}$ & {\vbox
to0pt{\vskip0pt\hbox to1in{\hfil$3584Ng^{N-8}$}\vfil}} \\
$\{\ell_1,\ell_2,\dots,\ell_6,N-1-\sum_{i=1}^{6}\ell_i\,;7\}$ & \\ \hline
$\{\ell_1,\ell_2,\dots,\ell_6,N-2-\sum_{i=1}^{6}\ell_i\,;7\}$ & $10752Ng^{N-9}$
\\ $\{\ell_1,\ell_2,\dots,\ell_7,N-2-\sum_{i=1}^{7}\ell_i\,;8\}$ &
$30720Ng^{N-10}$ \\ \hline
$\{\ell_1,\ell_2,\dots,\ell_7,N-3-\sum_{i=1}^{7}\ell_i\,;8\}$ & {\vbox
to0pt{\vskip0pt\hbox to1in{\hfil$84480Ng^{N-11}$}\vfil}}
 \\
$\{\ell_1,\ell_2,\dots,\ell_8,N-2-\sum_{i=1}^{8}\ell_i\,;9\}$ & \\ \hline
$\qquad\qquad\vdots$ & $\vdots$\qquad{} \\  
\end{tabular}
\end{center}   \medskip

From Table 3 one can immediately see that the first $\sigma=\{N;1\}$ gets
maximal $Q_\sigma$ if $g>4$.  The second boundary condition $\sigma=\{\ell,
N-\ell;2\}$ becomes maximal  if $4>g>3$.  In other words,
``all orientations organized into one wheel" becomes dominant for $g>4$, and
``$\ell$ and $ N-\ell$ orientation-modules  organized respectively into
independent wheels" is dominant if
$4>g>3$.  One can generally prove this fact including all lower power
$g^L$-terms as in the followings.

Let
$\sigma^*=\{\ell_1,\ell_2,\cdots,N-(N-L-m)-\sum_{i=1}^{m-1}\ell_i\,;m\}$
be a maximal configuration of $g^L$-terms (with $N-L-m\geq0$ being the number of
disintegrated modules from the wheels).  Then using equation
(\ref{combinatorial}) and $Q_{\sigma^*}\geq Q_\sigma$ for given $N,L$, one
gets   
\begin{equation}
   3m-2\geq 2(N-L)\geq 3m-5, \qquad N-L\geq m .
\label{m}
\end{equation}
This implies that under the change  $L\rightarrow L-1$ the $m$-value $m_L$
associated to the maximal configuration is restricted as
\begin{equation}
   m_L-\frac{1}{3}\leq m_{L-1}\leq m_L+\frac{5}{3}.
\end{equation}
Thus we have either $m_{L-1}=m_{L}$ or $m_{L}+1$.  In Table 3 both cases
are realized. At $L=N-5, N-8, N-11, \dots$ there arise two maximal
configurations $\sigma^*_1=
\{\ell_1,\ell_2,\cdots,N-(N-L-m)-\sum_{i=1}^{m-1}\ell_i\,;m\}$ and $\sigma^*_2=
\{\ell_1,\ell_2,\cdots,N-(N-L-m+1)-\sum_{i=1}^{m-2}\ell_i\,;m-1\}$  that have
the same maximal factor $Q_{\sigma^*_1}=Q_{\sigma^*_2}$.   This situation
always occurs for such $(N,L,m)$ that satisfies $2(N-L)=3m-5$.  (From the
restriction $N-L\geq m$, equation (\ref{m}), the case $L=N-2$ is excluded from
this argument.)   Choosing the appropriate configuration $\sigma^*(=\sigma^*_1 
\mbox{ or }\sigma^*_2)$ as the maximal configuration for $g^L$-term if
necessary, one can always consider $m_L$ as satisfying $m_{L-1}=m_L+1$. (See
Table 3.)

Now we consider the condition $Q_{\sigma^*}|_L > Q_{\sigma^*}|_{L-1}$ under
$m_{L-1}=m_L+1$ between neighboring maximal $\sigma^*$ at $g^L$- and
$g^{L-1}$-terms.  From the expression (\ref{combinatorial}) this condition is
recast into  
\begin{equation}
   g > \frac{2(N-L+1)}{m_L}, 
\end{equation}
for $L\leq N-2$.  However, the right-hand side of this condition is restricted
by the inequality (\ref{m}) as in
\begin{equation}
    3\geq \frac{2(N-L+1)}{m_L} \geq 3-\frac{3}{m_L}.
\end{equation}
Therefore, one concludes that as long as $g>3$ the relation $Q_{\sigma^*}|_L >
Q_{\sigma^*}|_{L-1}$ is always satisfied for all $L(\leq N-2)$, and the
configurations $\sigma=\{\ell, N-\ell;2\}$ or $\sigma=\{N;1\}$ get largest
combinatorial factors $Q_\sigma$ in all possible $\sigma$'s.  Obviously,
$\sigma=\{N;1\}$  is dominant if $g>4$, and so is $\sigma=\{\ell, N-\ell;2\}$ 
if $4>g(>3)$.  This completes the proof.

To summarize we have estimated the partition function (\ref{partition}) of our
model that consists of the sum of frequencies (or probabilities) $Q_\sigma$
for the occurrence of various (possible) organizations $\sigma$ of the
orientation-modules.  In the sum the organizations containing $M$ double-arcs
(Figure $9(b)$) has been shown to have extra ${\cal A}={\cal
A}_0e^{-\Delta{\cal E}_0/\lambda}$ factor $Q_\sigma\propto g^L{\cal A}^M$,
where ${\cal A}_0$, $\Delta{\cal E}_0$, $\lambda$ are the measures for
hypercolumn area in the unit of $\sigma_0^2$, average energy per orientation
module, and a coefficient in the lateral correlation $\Lambda_{\vr,\vr^*}$,
respectively.  We have suppressed those (higher order) terms by appropriate
choice of $\lambda$ so that the factor ${\cal A}<1$.  Among the remaining
non-double-arc organizations (Figure $9(c)$), we have proved the following
remarkable fact:  once $g(\equiv \zeta{\cal A})$ exceeds $3$, most probable
organizations are ``all orientation-modules organized into one
orientation-wheel (if $g>4$) or two wheels (if $g<4$)".  This is remarkable
since we do not need any fine-tuning of our parameter $\lambda$ into special,
definite value to obtain this result.   Since $\zeta(=$ measure of the
boundary length of the hypercolumn in the unit of $\sigma_0$) is roughly
equal to $24\sqrt{3}$ (Yamagishi, 1994), the condition $g>3$ holds under quite
natural circumstance.  In contrast, more than 4 orientation-wheels per
hypercolumn  are not easily  realized without fine-tuning of $g$ (or
$\lambda$-value).  Also remarkable is the fact that this result is independent
of the total module number $N$ as long as $N$ is large. 

\section{Summary}
In this paper we have considered the modular organization where the feature
space consists of infinite (continuous) degrees of freedom.  Specifically we
have considered the local organization of orientation modules while restricting
ourself to a hypercolumn area.  Since we do not know the minimal increment (or
resolution) of the orientation preference between neighboring modules at
low-level visual processing (V1), we proposed here statistical definition for
orientation-module boundaries.  Each module size in this definition was
considered indefinite and was defined only statistically.   Their size was
considered to be \underline{selected} in the course of higher-level visual
processing.  In the organization of such modules semi-local continuity of the
orientation preference angles was required for the scheme to work well.  In our
model of generalized Kohonen's feature mappings we have successfully realized
this scheme.  We have also shown in this model that the existence of
orientation-wheels observed in visual cortex is a consequence of the celebrated
Riemann-Hurwitz formula from (geometric) topology.

This formula from topology came into the game since in our model the
holomorphicity plays important role to determine the stationary points of the
net, and in two dimensions a holomorphic mapping is (locally) conformal. Hence
the topological property has turned out to play essential role.

Within the scope of the same topological argument we have discussed  the most
adequate number for orientation-wheels per hypercolumn.  We have estimated it
by  evaluating   the partition function ${\cal Z}_N$, equation
(\ref{partition}).  The resulting most probable orientation-wheel consists of 
all orientation modules organized monotonically with respect to orientation
angles if the lateral correlation strength $\lambda$ is chosen in such a way
that $g(=\zeta{\cal A}_0e^{-\Delta{\cal E}_0/\lambda}$, where $\zeta$
[$\approx24\sqrt{3}$ from the previous work I] being a measure of the
boundary-length of the hypercolumn region) exceeds 4 and ${\cal 
A}_0e^{-\Delta{\cal E}_0/\lambda}<1$.   Even if that is not the case, if $g$
exceeds 3, we have two wheels consisting of all orientations.  In contrast,
more than four orientation-wheels per hypercolumn are hard to realize
without fine-tuning of the parameter $g$.  This is  remarkable since  we have
the adequate organization of the orientation-wheels independently of the minimal
increment ($\approx\frac{180^\circ}{N}$) of the orientation angles.  This
number of orientation-wheels is also consistent with the observed result on
macaque monkeys.  This property remains true under the change of $N$ as long as
$N$ large without re-tuning the parameter $\lambda$ (hence
$g$-value)\footnote[2]{The most of the result of this paper is correct if
$N$ is larger than 10, approximately.}.

We have a couple of important issues left untouched in this paper. 
Basically what we have considered here are the things that could be handled
with tools only from topology.  However, for the analysis of the real cortical
surface, such as global organization of the orientation modules over the broad
range, we have to go beyond the topology.  This is very difficult but
important issue.  Another thing to be clarified is the selection mechanism
 of the  iso-orientation width (or increment) in the organized wheels that we
assumed as higher-cortical function.  We would like to return to these issues
in the future publications.

\bigskip\bigskip

Part of this work was performed under the auspices of the U.S. Department of
Energy by the Lawrence Livermore National Laboratory under Contract
W-7405-ENG-48 with the University of California.

\newpage

\noindent
{\Large\bf References}



\newpage

\noindent
{\Large\bf Figure Legends}



[Figures are embedded in the text.]

\end{document}